\documentclass[twocolumn,twocolappendix]{aastex7}

\newcommand{\RN}[1]{\MakeUppercase{\romannumeral #1}}

\usepackage{amsmath}
\shorttitle{Exploring the redshift distribution of Einstein Probe X-ray transients}
\shortauthors{O'Connor et al.}
\submitjournal{ApJL}
\usepackage{siunitx}
\usepackage{booktabs}
\usepackage{longtable}
\usepackage{caption}

\definecolor{blazeorange}{rgb}{1.0, 0.4, 0.0}
\definecolor{seagreen}{rgb}{0.18, 0.55, 0.34}
\definecolor{darkgreen}{rgb}{0.08, 0.45, 0.2}
\definecolor{rufous}{rgb}{0.66, 0.11, 0.03}
\definecolor{royalfuchsia}{rgb}{0.79, 0.17, 0.57}
\definecolor{scarlet}{rgb}{1.0, 0.13, 0.0}
\definecolor{royalpurple}{rgb}{0.47, 0.32, 0.66}

\usepackage{float}

\begin{document}

\title{
%Exploring the nature of Einstein Probe transients: clues from their redshift distribution \\
The redshift distribution of Einstein Probe transients supports their relation to gamma-ray bursts
}

\correspondingauthor{Brendan O'Connor}
\author[0000-0002-9700-0036]{Brendan O'Connor}
    \altaffiliation{McWilliams Fellow}
    \affiliation{McWilliams Center for Cosmology and Astrophysics, Department of Physics, Carnegie Mellon University, Pittsburgh, PA 15213, USA}
    \email[show]{boconno2@andrew.cmu.edu}

\author[0000-0001-7833-1043]{Paz Beniamini}
    \affiliation{Department of Natural Sciences, The Open University of Israel, P.O Box 808, Ra'anana 4353701, Israel}
    \affiliation{Astrophysics Research Center of the Open university (ARCO), The Open University of Israel, P.O Box 808, Ra’anana 4353701, Israel}
    \affiliation{Department of Physics, The George Washington University, Washington, DC 20052, USA}
    \email{paz.beniamini@gmail.com}
    
\author[0000-0002-1869-7817]{Eleonora Troja}
    \affiliation{Department of Physics, University of Rome ``Tor Vergata'', via della Ricerca Scientifica 1, I-00133 Rome, Italy}
    \email{eleonora.troja@uniroma2.it}

\author[0009-0001-0574-2332]{Malte Busmann}
    \altaffiliation{Recipient of a Wübben Stiftung Wissenschaft Student Grant}
    \affiliation{University Observatory, Faculty of Physics, Ludwig-Maximilians-Universität München, Scheinerstr. 1, 81679 Munich, Germany}
    \email{m.busmann@physik.lmu.de}
    
\author[0000-0001-6849-1270]{Simone Dichiara}
    \affiliation{Department of Astronomy and Astrophysics, The Pennsylvania State University, 525 Davey Lab, University Park, PA 16802, USA}
    \email{sbd5667@psu.edu}

\author[0000-0003-0516-2968]{Ramandeep Gill}
    \affiliation{Instituto de Radioastronom\'ia y Astrof\'isica, Universidad Nacional Aut\'onoma de M\'exico, Antigua Carretera a P\'atzcuaro $\#$ 8701,  Ex-Hda. San Jos\'e de la Huerta, Morelia, Michoac\'an, C.P. 58089, M\'exico }
    \affiliation{Astrophysics Research Center of the Open university (ARCO), The Open University of Israel, P.O Box 808, Ra’anana 4353701, Israel}
    \email{rsgill.rg@gmail.com}
    
\author[0000-0001-8530-8941]{Jonathan Granot}
    \affiliation{Department of Natural Sciences, The Open University of Israel, P.O Box 808, Ra'anana 4353701, Israel}
    \affiliation{Astrophysics Research Center of the Open university (ARCO), The Open University of Israel, P.O Box 808, Ra’anana 4353701, Israel}
    \affiliation{Department of Physics, The George Washington University, Washington, DC 20052, USA}
    \email{granot@openu.ac.il}

\author[0000-0002-1103-7082]{Michael J. Moss}
    \affiliation{NASA Postdoctoral Program Fellow, NASA Goddard Space Flight Center, Greenbelt, MD 20771, USA}
    \email{mikejmoss3@gmail.com}

\author[0000-0002-9364-5419]{Xander J. Hall}
	\affiliation{McWilliams Center for Cosmology and Astrophysics, Department of Physics, Carnegie Mellon University, Pittsburgh, PA 15213, USA}
	\email{xjh@andrew.cmu.edu}

\author[0000-0002-6011-0530]{Antonella Palmese}
	\affiliation{McWilliams Center for Cosmology and Astrophysics, Department of Physics, Carnegie Mellon University, Pittsburgh, PA 15213, USA}
	\email{palmese@cmu.edu}

%\author[0000-0002-0216-3415]{Rosa L. Becerra}
%    \affiliation{Dipartimento di Fisica, Universit\`a di Tor Vergata, Via della Ricerca Scientifica, 1, 00133 Rome, Italy}
%    \email{rosa.becerra@roma2.infn.it}

\author[0009-0007-6886-4082]{Niccol\`o Passaleva}
    \affiliation{Dipartimento di Fisica, Universit\`a di Tor Vergata, Via della Ricerca Scientifica, 1, 00133 Rome, Italy}
    \affiliation{Dipartimento di Fisica, Universit\`a di Roma “La Sapienza”, Ple Aldo Moro, 2, 00185 Rome, Italy}
    \email{niccolo.passaleva@uniroma1.it}

\author[0000-0003-0691-6688]{Yu-Han Yang}
    \affiliation{Dipartimento di Fisica, Universit\`a di Tor Vergata, Via della Ricerca Scientifica, 1, 00133 Rome, Italy}
    \email{yyang@roma2.infn.it}

%###############

\begin{abstract}
The launch of the \textit{Einstein Probe} unleashed a new era of high-energy transient discovery in the largely unexplored soft X-ray band. The \textit{Einstein Probe} has detected a significant number of fast X-ray transients that display no gamma-ray emission, complicating their robust association to more common gamma-ray bursts. To explore their possible connection, we analyzed the redshift distribution of both \textit{Einstein Probe} fast X-ray transients and long duration gamma-ray bursts. A comparative analysis of their cumulative redshift distributions using  non-parametric two-sample tests, namely the Kolmogorov-Smirnov and Anderson-Darling tests, finds no statistically significant difference. These tests favor that their redshifts are drawn from the same underlying distribution. This empirical connection between \textit{Einstein Probe} transients and long gamma-ray bursts is further supported by their agreement with the so-called ``Amati relation'' between the spectral peak energy and the isotropic-equivalent energy. Together, these results indicate that most extragalactic \textit{Einstein Probe} fast X-ray transients are closely related to long gamma-ray bursts and originate from a massive star (collapsar) progenitor channel. Our findings highlight the role of the \textit{Einstein Probe} in uncovering the missing population of failed jets and dirty fireballs that emit primarily at soft X-ray wavelengths. 
\end{abstract}



\section{Introduction} 

The high-energy transient sky has historically been dominated by long-duration gamma-ray bursts (GRBs; \citealt{Kouveliotou1993}), which are detected at an observed rate of approximately one per day. These powerful explosions are linked to the deaths of massive stars \citep[e.g., collapsars;][]{Woosley1993,MacFadyen1999}, which launch highly collimated, ultrarelativistic jets \citep{Frail1997}. The resulting emission is detectable across the electromagnetic spectrum, from gamma-rays to radio wavelengths \citep{Sari1998, Wijers1999,Granot2002}.

The launch of the \textit{Einstein probe} \citep[EP;][]{EP2015,EP2022,Yuan2025} has seen a sharp rise in the discovery rate of fast X-ray transients. The majority of these high-energy transients do not display any gamma-ray emission, while a handful have been solidly associated to GRBs \citep{Yin2024,Liu2024}. As such, the exact nature of the majority of the EP transient population, and their relation to the deaths of massive stars, is uncertain. While the broad class of fast X-ray transients display a great diversity in their properties \citep{Quirola2022,Quirola2023}, the fast transients discovered by EP, even those without prompt gamma-ray detections, tend to show GRB-like afterglow properties and energetics \citep{Yin2024,Liu2024,Sun2024,Ricci2025,Busmann2025,Yadav2025,Jiang2025}. Additionally, a number of nearby EP transients have been associated to Type Ic-BL supernova \citep{vanDalen2024,Srivastav2024,Rastinejad2025EP,Srinivasaragavan2025EP0108a}, typical of those found to follow long GRBs \citep{Woosley2006}. All together, there are multiple lines of evidence that strongly suggest that many EP transients are related to GRBs, or at least come from similar progenitors. 

In this work, we utilize the observed redshift distribution of \textit{Einstein Probe} detected fast X-ray transients to probe their connection to long duration gamma-ray bursts. The redshift distribution of a class of astrophysical transients imprints key information on their origins, including their connection to star formation. We find that properties of EP transients, including their overall redshifts and energetics, align closely with long GRBs, supporting their association to the core collapse deaths of massive stars. 

Throughout the manuscript we adopt a standard $\Lambda$CDM cosmology \citep{Planck2020} with $H_0$\,$=$\,$67.4$ km s$^{-1}$ Mpc$^{-1}$, $\Omega_\textrm{m}$\,$=$\,$0.315$, and $\Omega_\Lambda$\,$=$\,$0.685$.

\section{Source Sample and Observations}
\label{sec:obs}

\begin{deluxetable*}{lcccccc}
\tablecaption{Catalog of EP sources used in this work. The WXT properties including X-ray photon index $\Gamma_\textrm{WXT}$ and peak and time-averaged X-ray flux are reported. The X-ray fluxes are in the $0.5$\,$-$\,$4$ keV band. 
\label{tab:EPtab} %\textcolor{purple}{[Update table and analysis with new EP redshifts during revision -- do we add EP250702a and EP250827b. They do not super fit the criteria (well, really EP250827b doesn't with its like 5 days latency; but could discuss in latency section...) -- EP250908c is very tentative and should consider whether to include or not...]}
}

\tablehead{
\colhead{\textbf{Name}} & \colhead{\textbf{Redshift}} & \colhead{\textbf{GRB?}} & \colhead{$\mathbf{\Gamma_\textbf{\textrm{WXT}}}$} & \colhead{$\mathbf{F_\textbf{\textrm{WXT,avg}}}$} & \colhead{$\mathbf{F_\textbf{\textrm{WXT,peak}}}$} & \colhead{\textbf{References}} \\
\colhead{} & \colhead{} & \colhead{} & \colhead{} & \colhead{\textbf{erg/cm$^\mathbf{2}$/s}} & \colhead{\textbf{erg/cm$^\mathbf{2}$/s}} & \colhead{}
}
\startdata
EP240315a & 4.859 & GRB 240315C &  $1.4\pm0.1$ & $(5.3^{+1.0}_{-0.7})\times10^{-10}$ & $(4.6^{+0.8}_{-0.7})\times10^{-9}$ & (1,2,3)  \\[0.5mm]
EP240414a & 0.401 & -- & $3.1^{+0.7}_{-0.8}$ & $(6.5^{+1.3}_{-1.0})\times10^{-10}$ & $(2.2\pm0.7)\times10^{-9}$ & (4,5,6)  \\[0.5mm]
EP240801a & 1.673 & XRF 240801B &  $1.99\pm0.18$ & $(4.8\pm3.1)\times10^{-10}$ & $(1.2^{+0.6}_{-0.8})\times10^{-8}$ & (7,8,9,10) \\[0.5mm]
EP240804a & 3.662 & GRB 240804B &  $0.7^{+1.2}_{-0.4}$ & $(6.1^{+2.6}_{-1.6})\times10^{-10}$ & -- & (11,12,13) \\[0.5mm]
EP240806a & 2.818 & --& $2.6^{+1.2}_{-1.0}$ & $(1.9^{+1.8}_{-0.6})\times10^{-9}$ & -- & (15,16) \\[0.5mm]
EP241021a & 0.748 & --& $1.8\pm0.6$ & $(3.31^{+0.13}_{-0.09})\times10^{-10}$ & $1.0\times10^{-9}$ & (16,17,18) \\[0.5mm]
EP241030a\tablenotemark{a} & 1.411 & GRB 241030A & $2.5^{+0.8}_{-0.7}$ & $(7.5^{+3.0}_{-2.4})\times10^{-11}$ & -- & (19,20,21) \\[0.5mm] %This is the afterglow of GRB 241030A, but is still a WXT trigger so include it.
EP241107a & 0.456 & --& -- & -- & $4.2\times10^{-9}$ & (22,23) \\[0.5mm]
EP241113a & 1.53 & --& $1.3\pm0.2$ & $(5.57^{+1.26}_{-0.76})\times10^{-10}$ & -- & (24,25) \\[0.5mm]
EP241217a & 4.59 & --& $1.9^{+0.7}_{-0.6}$ & $(7.3\pm0.3)\times10^{-10}$ & -- & (26,27)\\[0.5mm]
EP241217b & 1.879 & GRB 241217A & $1.57^{+0.22}_{-0.21}$ & $(1.19\pm0.10)\times10^{-9}$ & -- & (28,29,30) \\[0.5mm]
EP250108a & 0.176 & -- &  $2.8\pm1.1$ & $(6.4^{+22.5}_{-3.0})\times10^{-11}$ & $(1.80^{+0.20}_{-0.06})\times10^{-10}$ & (31,32,33,34) \\[0.5mm]
EP250125a & 2.89 & -- & $0.8\pm0.5$ & $(1.8^{+0.7}_{-0.5})\times10^{-9}$ & -- & (35,36) \\[0.5mm]
EP250205a & 3.55 & GRB 250205A & $2.5^{+1.7}_{-1.2}$ & $(4.2\pm1.1)\times10^{-10}$ & -- & (37,38,39) \\[0.5mm]
EP250215a & 4.61 & GRB 250215A & -- & -- & -- & (40,41,42) \\[0.5mm]
EP250223a & 2.756 & --& $2.1\pm0.6$ & $(4.4^{+1.4}_{-1.1})\times10^{-10}$ & $2.0\times10^{-9}$ & (43,44)\\[0.5mm]
EP250226a & 3.315 & GRB 250226A & -- & -- & $9.8\times10^{-9}$ & (45,46,47) \\[0.5mm]
EP250302a & 1.131 & --& $0.6\pm0.4$ & $(7.0^{+2.0}_{-1.6})\times10^{-9}$ & $9.0\times10^{-9}$ & (48,49) \\[0.5mm]
EP250304a & 0.200 &-- & $2.2\pm0.1$ & $(5.3^{+0.4}_{-0.4})\times10^{-10}$ & -- & (50,51) \\[0.5mm]
EP250321a & 4.368 &-- & $0.66\pm0.17$ & $(1.73^{+0.21}_{-0.19})\times10^{-9}$ &  $4.2\times10^{-9}$ & (52,53) \\[0.5mm]
EP250404a & 1.88 & GRB 250404A & $0.41^{+0.26}_{-0.28}$ & $(5.9^{+2.2}_{-1.6})\times10^{-8}$ & -- & (54,55,56,57) \\[0.5mm]
EP250416a & 0.963 & GRB 250416C &  $0.32^{+1.00}_{-0.78}$ & $(5.66\pm1.77)\times10^{-9}$  & $(1.9\pm0.8)\times10^{-8}$  &  (58,59,60) \\[0.5mm]
EP250427a & 1.52 & GRB 250427A & $1.70^{+0.36}_{-0.34}$ & $(1.96^{+0.31}_{-0.23})\times10^{-9}$ & $2.0\times10^{-8}$ & (61,62,63,64) \\[0.5mm]
EP250704a\tablenotemark{b} & 0.661 & GRB 250704B & $1.7\pm1.3$ & $(1.3^{+0.8}_{-1.1})\times10^{-9}$  & $6.0\times10^{-7}$ & (65,66,67) \\[0.5mm]
EP250821a & 0.577 & -- &$1.2\pm0.5$ &$(1.3\pm0.4)\times10^{-9}$ & -- & (68,69,70) \\[0.5mm]
EP250827a & 1.61 & -- & $0.73^{+0.50}_{-0.47}$ & $(1.71^{+0.42}_{-0.34})\times10^{-9}$ & -- & (71,72,73) \\[0.5mm]
%\textcolor{red}{EP250908c} & 2.87 & -- &  &  &  & \\[0.5mm]
%\textcolor{red}{EP250911a} & 3.84 & -- &  &  &  & \\[0.5mm]
% &  & -- &  &  &  & \\[0.5mm]
% &  & -- &  &  &  & \\[0.5mm]
% &  & -- &  &  &  & \\[0.5mm]
% &  & -- &  &  &  & \\[0.5mm]
% &  & -- &  &  &  & \\[0.5mm]
% &  & -- &  &  &  & \\[0.5mm]
% &  & -- &  &  &  & \\[0.5mm]
% &  & -- &  &  &  & \\[0.5mm]
% &  & -- &  &  &  & \\[0.5mm]
% &  & -- &  &  &  & \\[0.5mm]
\enddata
%\tablecomments{text}
\tablenotetext{a}{EP241030a is likely the afterglow of GRB 241030A and not the prompt emission \citep{241030a}. }
\tablenotetext{b}{EP250704a is associated to the short duration GRB 250704B \citep{250704a-grb}. }
\tablerefs{ (1) \citet{Liu2024}, (2) \citet{Levan2024}, (3) \citet{EP240315a-z}, (4) \citet{Sun2024}, (5) \citet{vanDalen2024}, (6) \citet{Srivastav2024}, (7) \citet{Jiang2025}, (8) \citet{EP240801a}, (9) \citet{240801a-z1}, (10) \citet{240801a-z2}, (11) \citet{240804a}, (12) \citet{240804a-grb}, (13) \citet{240804a-z}, (14) \citet{240806a}, (15) \citet{240806a-z}, (16) \citet{Shu2025}, (17) \citet{GTCz}, (18) \citet{VLTz}, (19) \citet{241030a} , (20) \citet{2410310a-grb}, (21) \citet{241030a-z}, (22) \citet{241107a}, (23) \citet{241107a-z}, (24) \citet{241113a}, (25) \citet{241113a-z}, (26) \citet{241217a}, (27) \citet{241217a-z}, (28) \citet{241217b}, (29) \citet{241217b-grb}, (30) \citet{241217b-z}, (31) \citet{Li2025}, (32) \citet{Rastinejad2025EP}, (33) \citet{Eyles-Ferris2025EP}, (34) \citet{Srinivasaragavan2025}, (35) \citet{EP250125a}, (36) \citet{EP250125a-z}, (37) \citet{250205a}, (38) \citet{250205a-grb}, (39) \citet{250205a-z}, (40) \citet{2502515a}, (41) \citet{250215a-grb}, (42) \citet{250215a-z}, (43) \citet{250223a}, (44) \citet{250223a-z}, (45) \citet{250226a}, (46) \citet{250226a-grb}, (47) \citet{250226a-z}, (48) \citet{2025GCN.39556....1D}, (49) \citet{0302a-vlt-redshift}, (50) \citet{250304a}, (51) \citet{250304a-z}, (52) \citet{250321a}, (53) \citet{250321a-z}, (54) \citet{250404a}, (55) \citet{250404a-grb}, (56) \citet{Yin2025ep250404a}, (57) \citet{250404a-z2}, (58) \citet{250416a}, (59) \citet{250416a-grb}, (60) \citet{250416a-z}, (61) \citet{250427a}, (62) \citet{250427a-grb}, (63) \citet{250427a-z1}, (64) \citet{250427a-z2}, (65) \citet{250704a}, (66) \citet{250704a-grb}, (67) \citet{250704a-z}, (68) \citet{250821a}, (69) \citet{250821a-z}, (70) \citet{250821a-z2}, (71) \citet{250827a}, (72) \citet{250827a-z}, (73) \citet{250827a-z2}
}
\end{deluxetable*}

\subsection{Sample of Einstein Probe Transients}
\label{sec:EP}

The \textit{Einstein probe} (EP) is a new soft X-ray mission \citep{EP2015,EP2022,Yuan2025} with wide-field capabilities. EP was launched on January 9, 2024, and is currently surveying the sky in the soft X-ray band between $0.5$\,$-$\,$4.0$ keV. The Wide-field X-ray Telescope (WXT) has an instantaneous field-of-view (FOV) of 3,600 deg$^2$, and is capable of autonomously detecting transients on-board the spacecraft \citep{Yuan2025,Zhao2025}. These EP/WXT transients are then generally reported through General Coordinate Network (GCN) Notices\footnote{\url{https://gcn.nasa.gov/notices}}, and, later, GCN Circulars\footnote{\url{https://gcn.nasa.gov/circulars}}. The delay of these reports from the EP/WXT trigger time varies from event to event based on the data's downlink latency, among other factors.

Due to the low energy range ($0.5$\,$-$\,$4.0$ keV) observed by WXT, flaring stars are a major source of contamination that must be filtered out to select a clean extragalactic sample of EP transients. 
%For example, between November 2024 and July 2025, 35 out of 54 EP transients that were reported in near real-time ($\leq$\,$4$ hr latency; Jamie Kennea, private communication) and followed-up through Priority 0 \citep[P0; see][]{Tohuvavohu2024} Target of Opportunity (ToO) observations\footnote{\url{https://www.swift.ac.uk/EP/}} by the \textit{Neil Gehrels Swift Observatory} \citep{Gehrels2004} X-ray Telescope \citep[XRT;][]{Burrows2005} turned out to be flaring stars. 
These contaminant sources are not provided a standard EP designation (i.e., EPYYMMDDa), and instead labeled by their trigger identification number (i.e., EP\#XXXXXXXXXXX). In order to avoid further contamination from Galactic sources, EP/WXT's on-board trigger algorithm does not relay alerts for transients lying at close proximity to the Galactic Plane (e.g., EP250702a; \citealt{250702a}). 

After removing these contaminating sources, % and relaxing the $\leq$\,$4$ hr latency requirement for P0 \textit{Swift} observations, 
the rate of extragalactic EP transients is $\sim$\,$70$\,$-$\,$80$ per year. As of August 29, 2025, we find $\sim$\,113 publicly reported EP/WXT transients fitting this criteria. While the information publicly reported by EP varies from event to event, in general the available quantities communicated based on the WXT trigger are: source localization, approximate duration, soft X-ray photon index, time-averaged (unabsorbed) flux, and, in some cases, the peak X-ray flux. Usually there is additional information based on observations with the Follow-up X-ray Telescope (FXT; \citealt{Chen2025fxt,Zhang2025fxt}), if it is available. In Table \ref{tab:EPtab}, we compile the available photon index and flux information on a per event basis for events in our sample, which is further described below.

\subsection{Compilation of the Einstein Probe Redshift Distribution}
\label{sec:redshifts}

We have compiled a sample of redshifts for EP/WXT transients  reported publicly in GCN Circulars or available in the published literature (including the spectra reported in \S \ref{sec:optspec}; Figure \ref{fig:speczs}). Our sample comprises all publicly reported EP sources up to August 29, 2025. In total, we find $\sim$\,113 publicly reported EP/WXT transients. Of these EP sources, we find 26 with secure spectroscopic redshifts. The redshift completeness is therefore only $\sim$\,$23\%$. The catalog of EP sources with redshifts is tabulated in Table \ref{tab:EPtab}. %The cumulative distribution function (CDF) of the EP transient redshift distribution is shown in Figure \ref{fig:redshift-CDFs}.

The majority of these redshifts are determined through the identification of absorption lines detected in optical spectroscopy, with the exception of a handful that are based on their association to a host galaxy (from the detection of underlying emission lines). The measure of absorption lines provides a robust lower limit to the redshift, which is generally assumed to be the precise redshift of the transient. While the detection of emission lines provides an accurate redshift for a given galaxy, there is always the possibility that the galaxy association is due to a (either foreground or background) chance alignment \citep{Bloom2002}. As such, these handful of EP transient redshifts are less secure (see Appendix \ref{caution} for further discussion).

\begin{figure*}
    \centering
\includegraphics[width=1.05\columnwidth]{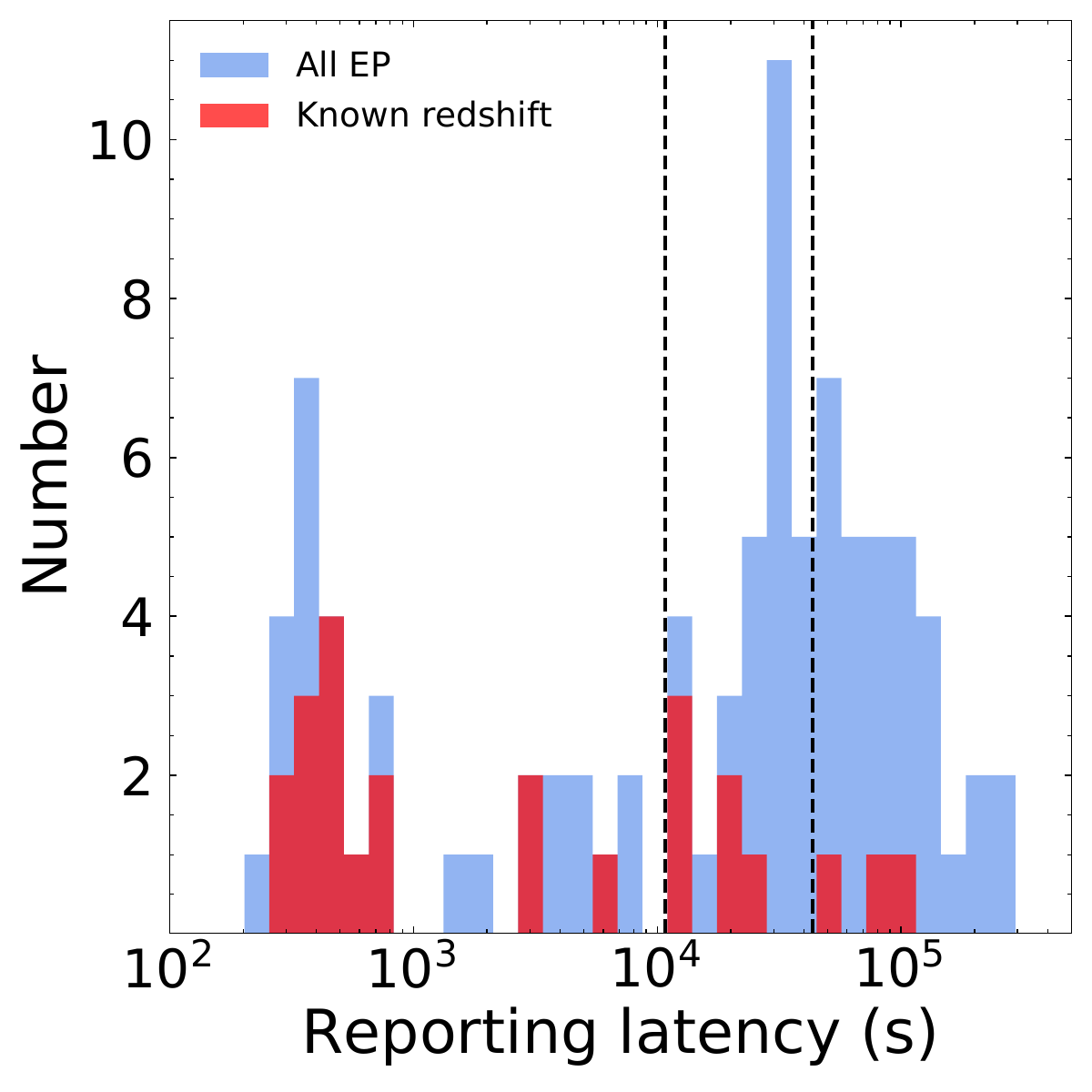}
\includegraphics[width=1.05\columnwidth]{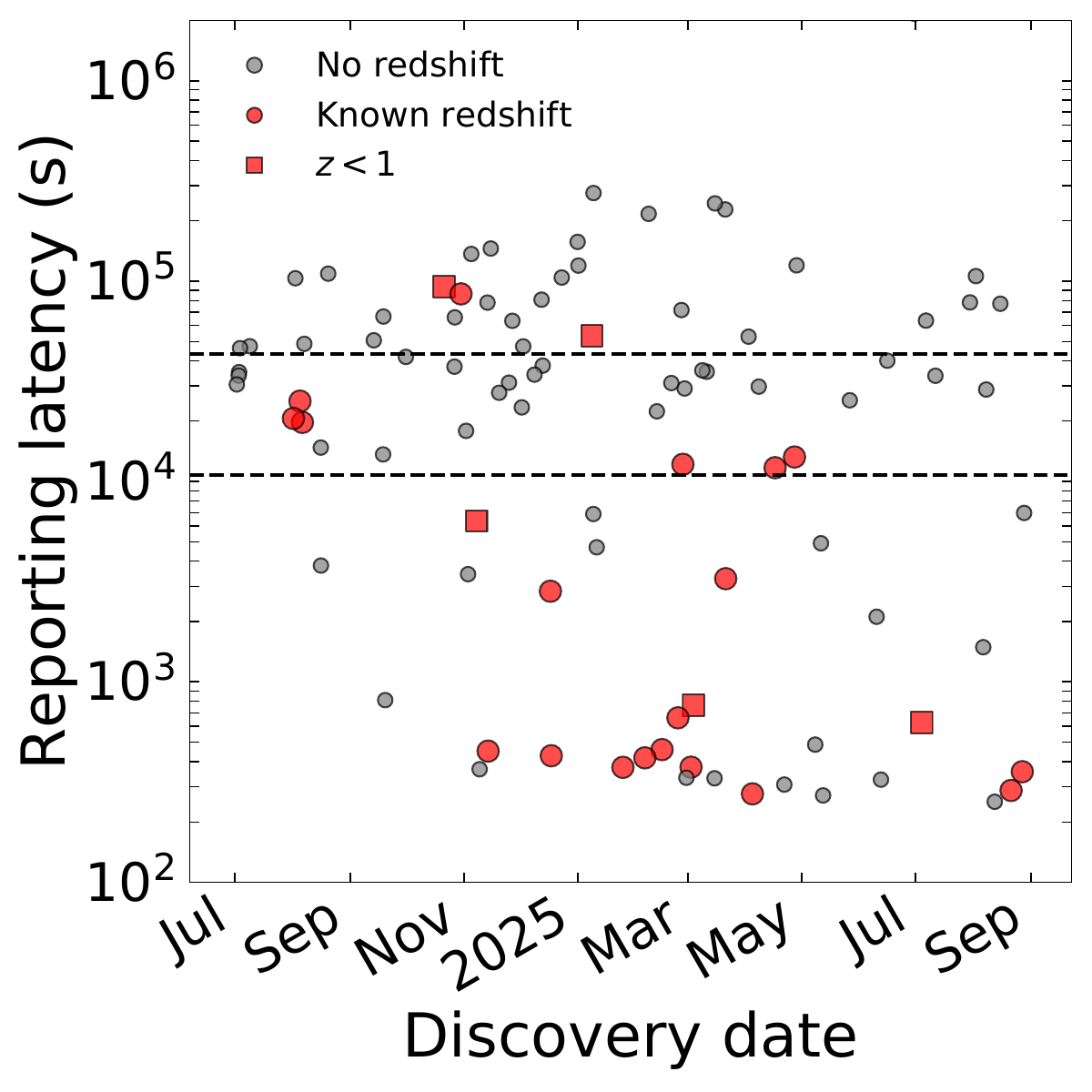}
    \caption{\textbf{Left:} Histogram of the EP reporting latency measured as the time between GCN notice and EP trigger time. All EP transients (from July 2024 onwards) are shown in blue, and those with measured redshifts marked in red. The dashed lines mark $3$ and $12$ hours from the EP trigger. \textbf{Right:} Reporting latency versus discovery date for EP transients. Those with no redshift are marked in gray, and those with redshifts in red. Sourced with redshift at $z$\,$<$\,$1$ are shown by squares.
    }
    \label{fig:delay}
\end{figure*}

\subsection{Redshift and Optical Afterglow Completeness}
\label{sec:complete}

The redshift completeness for EP transients is lower than the completeness for \textit{Swift} GRBs \citep[e.g.,][]{Hjorth2012,Jakobsson2012,Perley2016,Selsing2019}.
However, whereas \textit{Swift} is equipped with the Ultra-Violet Optical Telescope (UVOT; \citealt{Roming2005}) for rapid optical follow-up, EP has to rely on ground-based follow-up, which introduces a delay in the discovery and characterization of the optical counterpart.

Here we discuss the impact of latency on the optical afterglow and redshift completeness. We retrieved the GCN Notice\footnote{A GCN Notice is a low-latency, machine readable alert, and differs from human readable GCN circulars that are generally sent on longer timescales across all ingested missions.} time and EP trigger time from the Astro-COLIBRI\footnote{\url{https://astro-colibri.com/}} platform \citep{2021ApJS..256....5R}. These values are available starting in July 2024 for 93 EP transients, which misses only the first $4$\,$-$\,$5$ months of the commissioning phase of EP. We therefore only lack this information for 18 out of 113 EP sources. Only 2 out of these 18 sources have redshifts (EP240315a and EP240414a). Therefore, we consider the completeness of EP transients using the 24 redshifts (out of 93\footnote{We note that this does not include the handful of EP events reported through GCN Circulars (and not reported through any GCN Notice) with longer delays (e.g., EP250827b) and only includes those with a GCN Notice.} events) uncovered after July 2024, which yields a similar redshift completeness of $\sim$\,$25\%$. Figure \ref{fig:delay} shows the reporting latency (defined as the delay between the on-board trigger and public notice to the community) of EP transients. 

For those reported within 12 hours, we find that 21 out of 60 have a measured redshift ($\sim$\,$35\%$). Shortening the delay to those reported through low-latency GCN Notice within 3 hours, we find 15 out of 32, i.e., $\sim$\,$47\%$, have a measured redshift, nearly a factor of 2 increase to the full sample when not accounting for reporting delay. Of the 26 (27\% of EP transients) reported publicly within 1 hour of discovery, 13 (50\%) have a measured redshift. 

This shows that reporting latency is an important factor in the discovery of optical afterglows \citep[see also][]{EP-lulin}, and therefore spectroscopic redshift measurements.
%, as already clearly demonstrated during the \textit{Swift} era of GRBs. 
We suggest that this delay has a major (artificial) impact on the optical afterglow completeness, and is large compared to \textit{Swift} GRBs. This discrepancy in completeness is due to the combination of a few factors, including \textit{i}) the shorter delay ($<$\,$1$ min) in public notice for \textit{Swift}\footnote{\url{https://gcn.nasa.gov/missions/swift}}, \textit{ii}) the shorter delay in acquiring arcsecond precision positions with the \textit{Neil Gehrels Swift Observatory} \citep{Gehrels2004} X-ray Telescope \cite[XRT;][]{Burrows2005}, and \textit{iii}) the lack of initial imaging with \textit{Swift} UVOT %Ultra-Violet Optical Telescope (UVOT; \citealt{Roming2005}) 
that allows for rapid spectroscopy with a knowledge of the optical brightness. While the prompt gamma-ray localizations of the \textit{Swift} Burst Alert Telescope \citep[BAT;][]{Barthelmy2005} are comparable to the WXT localization, the XRT and UVOT localizations are significantly more accurate and avoid necessary image scanning over a multi-arcminute field. This aids in the rapid identification of the optical counterpart, increasing the likelihood of spectroscopy \citep[e.g.,][]{Selsing2019}. 
We note that \textit{Swift} performs rapid Priority 0 \citep[P0; see][]{Tohuvavohu2024} Target of Opportunity (ToO) observations\footnote{\url{https://www.swift.ac.uk/EP/}} for EP transients reported in near real-time ($\leq$\,$4$ hr latency; Jamie Kennea, private communication), which can provide a more rapid, precise localization of the X-ray source. This may be a factor in the increasing redshift completeness for EP transients reported in lower latency. 

As there is no intrinsic difference in a source with a longer or shorter latency in reporting its discovery, we consider that this does not have a large impact on our results or interpretation. In fact, it supports our overall conclusion by showing that the low redshift completeness of EP transients is not necessarily an intrinsic property. While it has been proposed that some EP fast X-ray transients are related to ``dark'' GRBs \citep{vanderhorst2009} based on their faint optical brightness at a few days after detection \citep{EP-lulin}, a comprehensive analysis is required to determine if this fraction of possible ``dark'' events is consistent with the fraction observed in GRBs or substantially different. A full investigation of optical darkness in the EP transient population and the optical afterglow completeness of EP fast X-ray transients will be presented in M. Busmann et al. (in preparation).

\subsection{Prompt Gamma-ray Properties}
\label{sec:gammalimits}

We can break down the sample of EP transients into two additional classes, which we consider separately throughout this manuscript. The first is EP-GRBs, being those with gamma-ray detections from other high-energy monitors. The second are EP transients without gamma-ray detections, which can be referred to as gamma-ray dark events \citep{Yadav2025}. As such, there are three samples of EP events (all those with redshift, those with redshift and gamma-rays, and those with redshift and without gamma-rays) that we consider separately in the analysis presented in \S \ref{sec:results}.

Of the sample of EP sources with redshifts (Table \ref{tab:EPtab}), 12 events (46\% of the sample) were also detected as gamma-ray bursts (referred to as EP-GRBs throughout this work). The coincident gamma-ray emission detected from these sources display long durations ($>$\,$2$ s), with the exception of the short burst EP250704a/GRB 250704B \citep{250704a,250704a-grb}. While the majority of EP-GRBs triggered high-energy gamma-ray monitors as standard gamma-ray bursts, some were subthreshold detections based on targeted searches. These include EP240315a/GRB 240315C \citep{Liu2024}, EP240801a/XRF 240801B \citep{Jiang2025}, and EP250427a/GRB 250427A \citep{250427a-grb}. In the case of EP241217b/GRB 241217A, while the gamma-ray emission was first detected \citep{241217b-grb} by the \textit{Space-based multi-band astronomical Variable Objects Monitor} \citep[\textit{SVOM};][]{Wei2016-SVOM}, a subthreshold detection \citep{241217b-grb2} was made by the \textit{Fermi Gamma-ray Space Telescope} Gamma-ray Burst Monitor \citep[GBM;][]{Meegan2009}. This highlights that limited gamma-ray sensitivity can be a factor in detecting the prompt emission.

While in the majority of cases EP has clearly detected the prompt emission of the transient, we note that EP241030a is very likely a detection of the afterglow of GRB 241030A and not the prompt emission \citep{241030a}. However, as this still represents an on-board EP/WXT detection (independent of the GRB trigger), we include this transient in our sample of redshifts. This may also be the case for EP250205a/GRB 250205A \citep{250205a-grb}, as EP detected the emission 410 s (90 s in rest frame) after the gamma-ray trigger due to Earth occultation \citep{250205a}. However, a full analysis of the available data is required to robustly make this determination on a case-by-case basis \citep[as for, e.g., EP250404a/GRB 250404A;][]{Yin2025ep250404a}, especially as many joint EP-GRB detections have demonstrated significantly longer duration soft X-ray emission \citep[see, e.g.,][]{Yin2024,Liu2024,Yin2025ep250404a}.

\begin{figure}
    \centering
\includegraphics[width=\columnwidth]{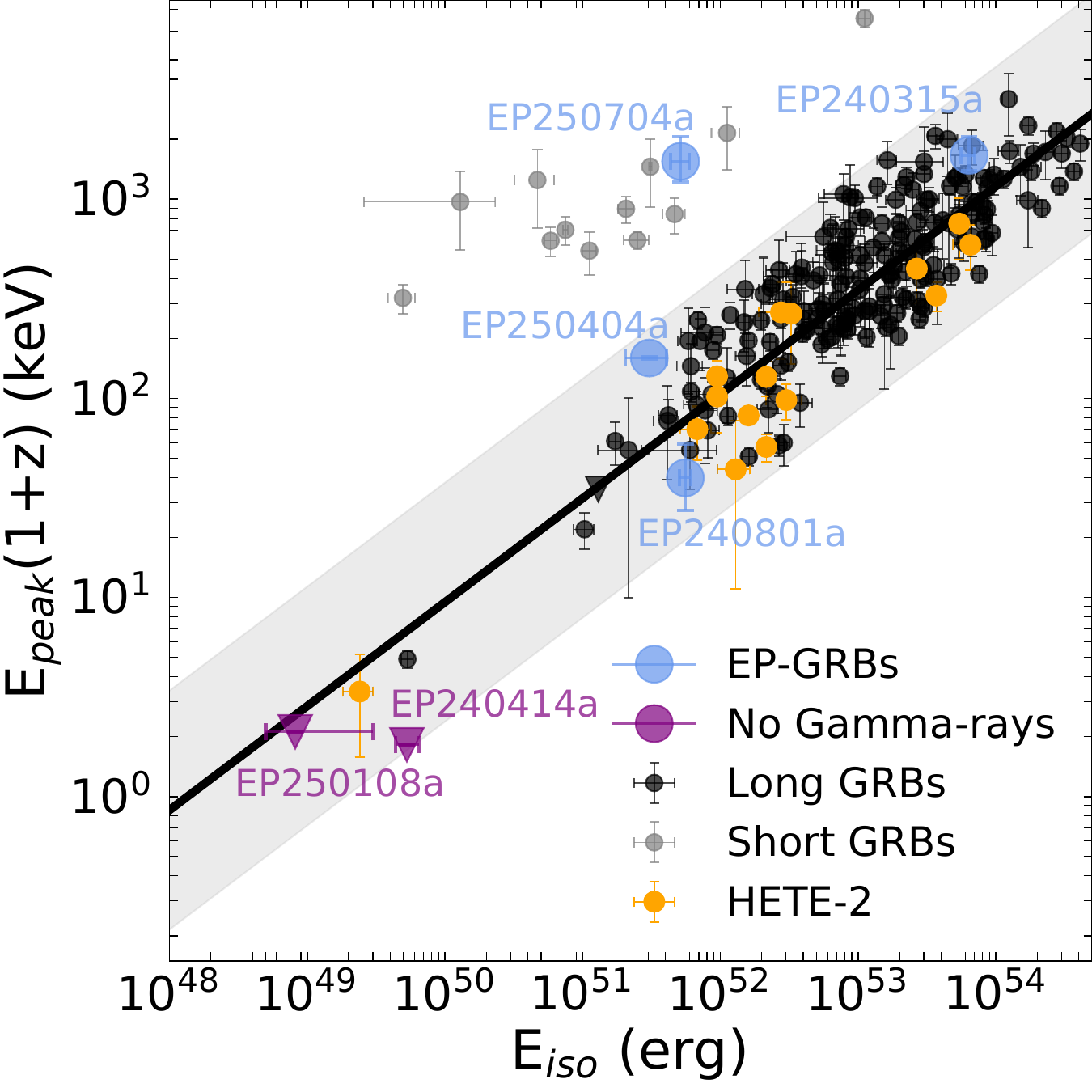}
    \caption{The $E_\textrm{p}$-$E_\textrm{iso}$ plane of GRBs with short GRBs shown in gray and long GRBs in black. EP transients with gamma-ray detections are shown in blue, and those without gamma-rays in purple with upper limits represented by downward triangles. We explicitly highlight \textit{HETE-2} bursts in orange \citep{Pelangeon2008}. The solid black line and $3\sigma$ scatter to the correlation for long GRBs are reproduced from \citet{Amati2019}. The figure is reproduced from \citet{Dichiara2021EE}. 
    %\textcolor{purple}{[Update if any new EP-z with Epeak.]}
    }
    \label{fig:Ep-Eiso}
\end{figure}

\begin{figure*}
    \centering
\includegraphics[width=1.05\columnwidth]{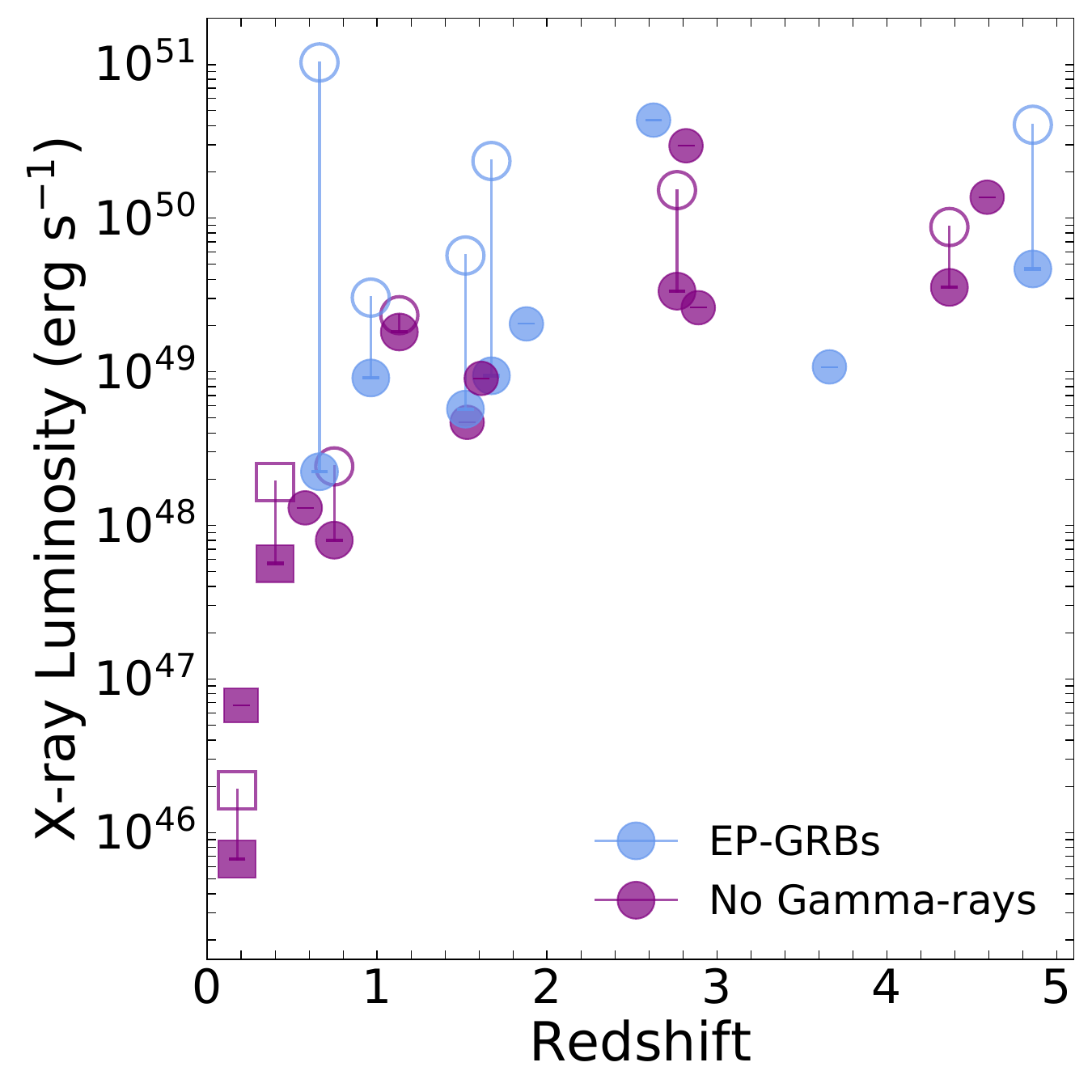}
\includegraphics[width=1.05\columnwidth]{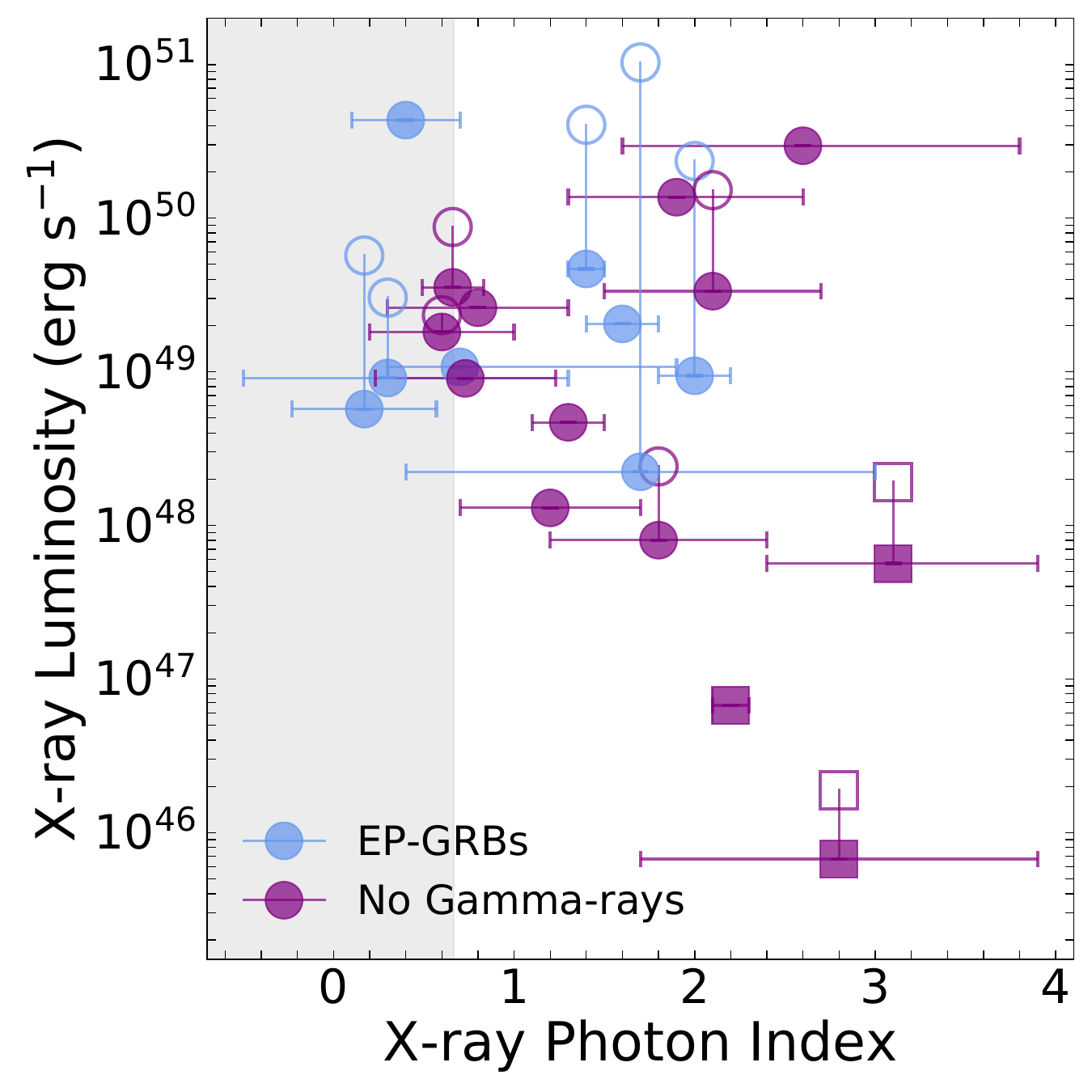}
    \caption{\textbf{Left:} Rest frame $0.5$\,$-$\,$4$ keV X-ray luminosity measured by EP/WXT versus redshift (Table \ref{tab:EPtab}). The time-averaged luminosity is represented by solid symbols and empty symbols refer to the peak luminosity. 
    We have designated between EP-GRBs (blue) and those without gamma-rays (purple). Sources represented by squares have been found to be associated to Type Ic-BL supernovae at $z$\,$\lesssim$\,$0.4$ \citep{vanDalen2024,Rastinejad2025EP,Srinivasaragavan2025EP0108a,EP250304a-SN-GCN}.  
    \textbf{Right:} Rest frame luminosity versus X-ray photon index measured by EP/WXT. The gray shaded region represents the synchrotron ``line of death'' \citep{Preece1998}. %\textcolor{purple}{[Update based on any new EP-z; starting with 250911a]}
    }
    \label{fig:lum}
\end{figure*}

\section{Analysis and Results}
\label{sec:results}

\subsection{High-energy Properties}
\label{sec:high-energy-properties}

While EP/WXT transients clearly display prompt X-ray emission, the majority do not trigger other high-energy hard X-ray or gamma-ray monitors. This can be due to a combination of the low peak energy of their prompt emission spectrum \citep[e.g., EP240414a, EP240801a, EP250108a;][]{Liu2024,Jiang2025,Li2025}, or the worse sensitivity of gamma-ray detectors compared to EP. Some events that do not trigger on-board the gamma-ray spacecraft have been identified as GRBs later in subthreshold searches, such as EP240219a/GRB 240219A \citep{Yin2024}, EP240315a/GRB 240315C \citep{Liu2024}, and EP240801a/XRF 240801B \citep{Jiang2025}. It is therefore possible that other EP sources also emit faint gamma-rays that cannot be detected to the available gamma-ray sensitivity limits (usually $10$\,$-$\,$100$ times less sensitive than EP/WXT). In Figure \ref{fig:Ep-Eiso}, we show the standard GRB correlation \citep[e.g.,][]{Amati2002,Amati2006} between the rest frame peak energy $E_\textrm{p}(1+z)$ and the isotropic equivalent gamma-ray energy $E_\textrm{iso}$ in the $1\,$-$10,000$ keV energy range. The sample of long GRBs shown is the same used in \citet{Amati2008,Amati2009}, and spans multiple missions including \textit{Swift}, \textit{Fermi}, \textit{Konus-Wind}, \textit{BATSE}, \textit{BeppoSAX}, and \textit{HETE-2}. It has been updated to include the latest results from \textit{Fermi} \citep{fermicat} and \textit{Konus-Wind} \citep{kwcat}. While only a handful of EP-GRBs have peak energies publicly reported in the literature \citep{Liu2024,Li2025,Jiang2025,Yin2025ep250404a}, those that do clearly follow the $E_\textrm{p}$-$E_\textrm{iso}$ plane and show an extension at the faint end of the $E_\textrm{iso}$ distribution. This is similar to the conclusion drawn from early intrinsic X-ray flash detections (i.e., XRF 020903) by \textit{HETE-2} \citep{Sakamoto2004}. The clear exception for EP-GRBs in Figure \ref{fig:Ep-Eiso} is the location of EP250704a/GRB 250704B \citep{250704a-grb}, which combined with its short gamma-ray duration ($\ll$\,$2$ s), strongly suggest that it is a true short duration GRB.

%[I checked and at least for THIS sample with redshifts it is not the case -- they have all prompt gamma-ray coverage.]
%An alternative is that the EP transient's position is not covered by gamma-ray satellites during the prompt phase. For example, \citet{Yadav2025} analyzed 84 EP events observed up to March 21, 2025. \citet{Yadav2025} found that 55 events had prompt emission coverage by the \textit{Fermi Gamma-ray Space Telescope} \citep{Meegan2009}, whereas 29 events were either Earth occulted or occurred during the South Atlantic Anomaly (SAA) passage. Additional searches have been performed to search for gamma-ray emission from other high-energy satellites \citep[e.g.,][]{Zhang2025-EP-GECAM-Search,RapidGBM}. In many cases where other spacecraft, such as \textit{Swift} or \textit{Fermi}, do not have available prompt emission coverage, \textit{Konus-Wind} \citep{Aptekar1995} can still provide limits on the prompt emission \citep[e.g.,][]{OConnor2025}, or, in some cases, a detection \citep[e.g.,][]{Liu2024}.

In Figure \ref{fig:lum}, we show the rest frame X-ray luminosity ($0.5$\,$-$\,$4$ keV) measured by EP/WXT versus redshifts (left panel) and ($0.5$\,$-$\,$4$ keV) X-ray photon index (right). The high luminosity of these sources out to $z$\,$\sim$\,$5$ is similar to those observed from gamma-ray bursts, albeit at different energies. In the right panel of Figure \ref{fig:lum}, we observe a weak trend that the lower redshift EP events, with intrinsically lower luminosities, have softer X-ray photon indices. This can be explained if their prompt emission spectral peak energies $E_\textrm{p}$ are low, as was the case for EP240414a \citep{Liu2024} and EP250108a \citep{Li2025}, which are shown in the left panel of Figure \ref{fig:lum} at $z$\,$=$\,$0.4$ and $0.176$, respectively. In this case, the prompt emission emits the bulk of its energy below the gamma-ray band, which can explain the lack of prompt gamma-ray detections. However, we do not only observe EP events without gamma-rays to have a steep photon index, and there is a clustering of three events around the synchrotron ``line of death'' \citep{Preece1998}.

%We further compiled the prompt gamma-ray properties of this sample of events from \citet{Yadav2025} and \citet{Zhang2025-EP-GECAM-Search}, and supplemented using public GCN Circulars where necessary. All events in our sample, including those without gamma-ray detections, have gamma-ray coverage of the EP/WXT prompt emission trigger time. For events without a prompt gamma-ray detection, we compiled $3\sigma$ upper limits to their gamma-ray flux in the $10$\,$-$\,$10,000$ keV band \citep[see][]{Yadav2025,Zhang2025-EP-GECAM-Search}. From Figure \ref{fig:lum}, it is clear that the typical gamma-ray sensitivity is significantly less sensitive than the detection threshold for EP/WXT.
%\textcolor{purple}{[Check whether I plotted the peak gamma-ray flux or time-averaged]}
%\textcolor{purple}{[Need to check the statements that ``all have prompt coverage'' as some of this sample isn't in the csv file I compiled -- should also say above which are excluded from the X-ray Lum plots... e.g., 250427A fermi subthreshold did not report flux or fluence information.]} 

\begin{figure*}
    \centering
\includegraphics[width=1.05\columnwidth]{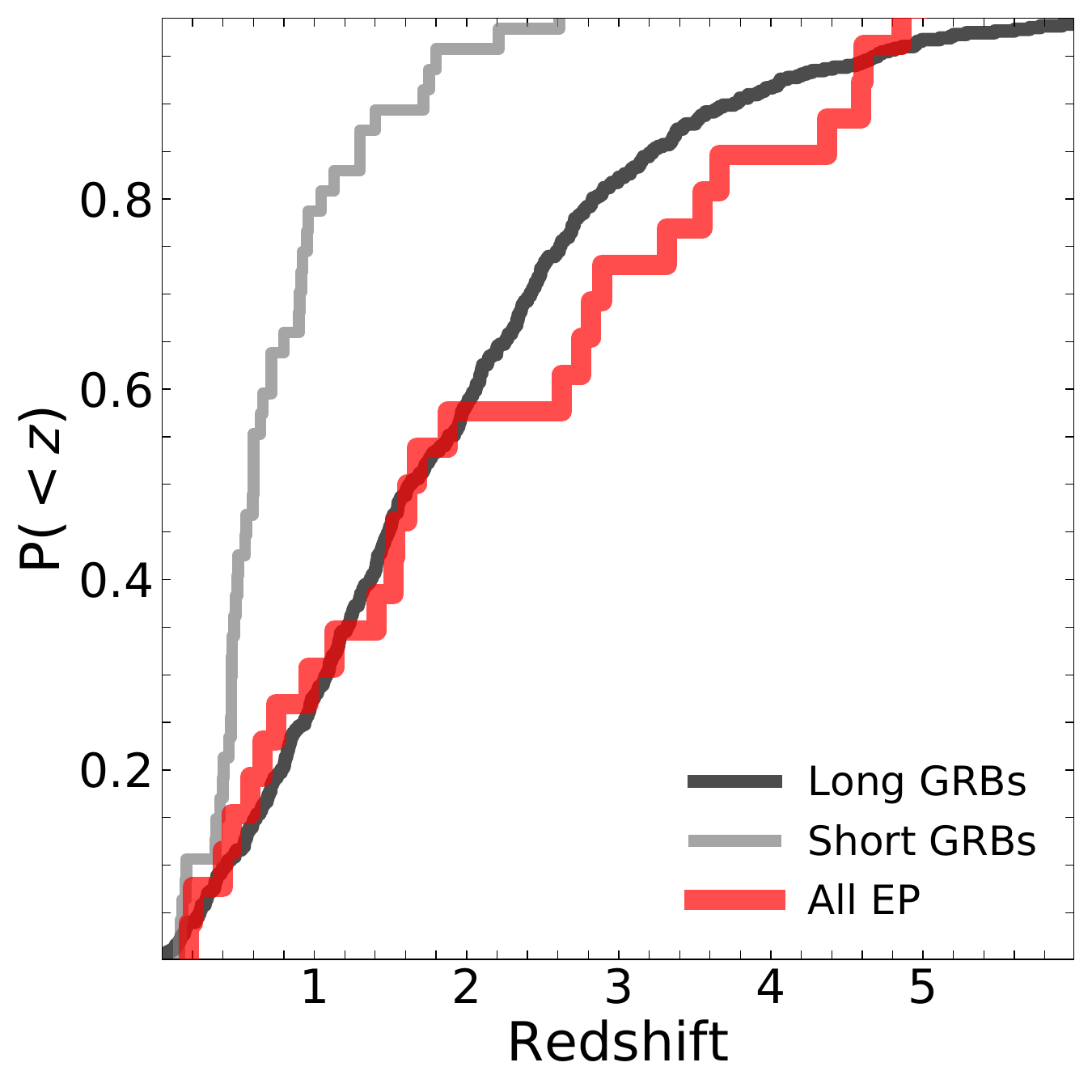}
\includegraphics[width=1.05\columnwidth]{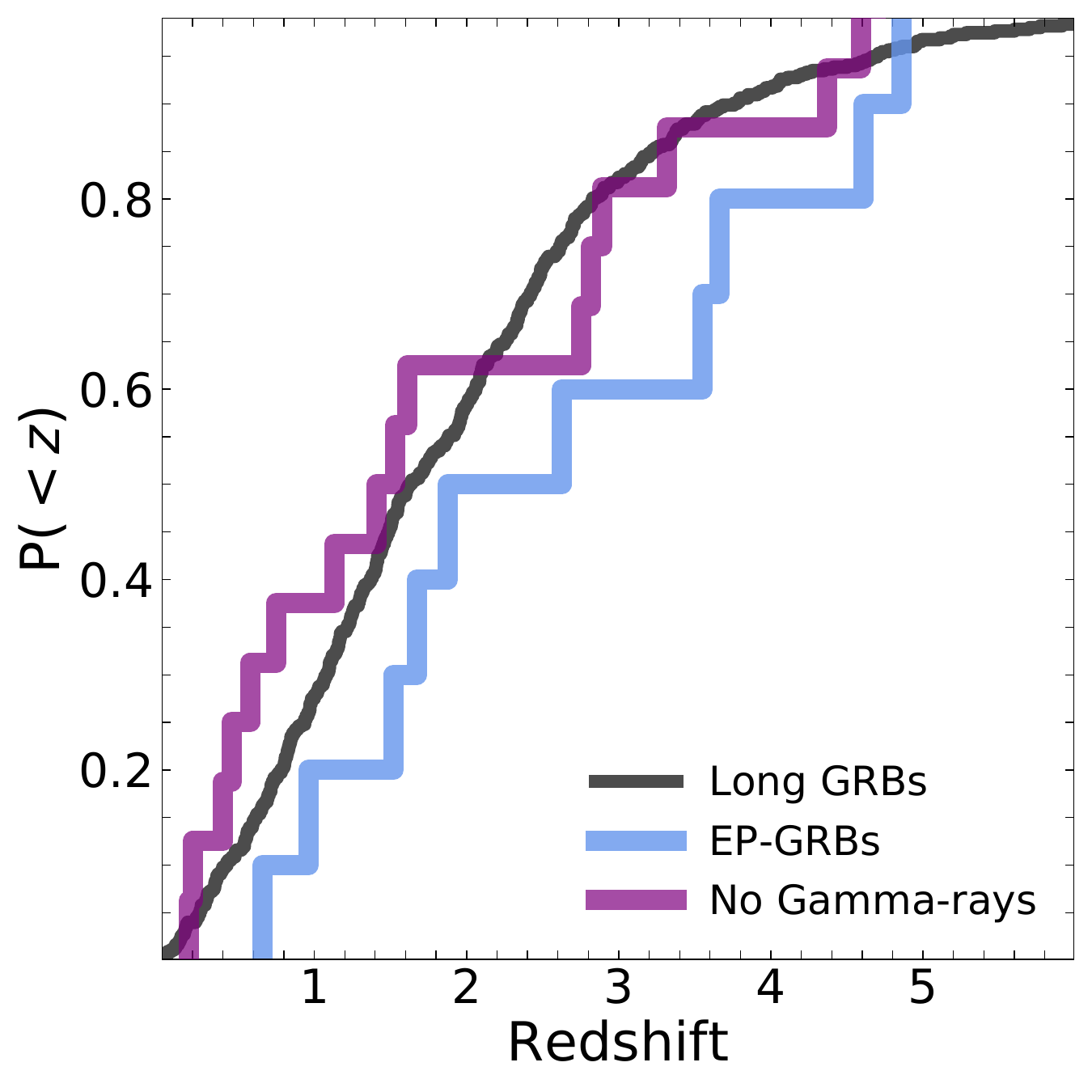}
    \caption{\textbf{Left:} Cumulative distribution of the redshift of EP transients (red) compared to short (gray; \citealt{OConnor2022}) and long (black; Greiner Catalog) GRBs. \textbf{Right:} The EP transients are further divided into those without gamma-ray detections (purple) and those with joint GRB detections (EP-GRBs; blue).
    }
    \label{fig:redshift-CDFs}
\end{figure*}

\subsection{Redshift Distribution Comparison}
\label{sec:bootstrap}

We aim to compare the redshift distribution of EP transients to long duration GRBs to probe the their relationship (Figure \ref{sec:redshifts}). We compare our compiled EP redshifts (Table \ref{tab:EPtab}) to the long GRB distribution compiled by the Greiner GRB Catalog\footnote{\url{https://www.mpe.mpg.de/~jcg/grbgen.html}}. The cumulative distribution function (CDF) of their redshift distributions is shown in the left panel of Figure \ref{fig:redshift-CDFs}, where we also show the distribution for the two sub-samples with and without gamma-ray detections (Figure \ref{fig:redshift-CDFs}; right panel). To statistically compare these populations, we applied non-parametric two-sample tests on the cumulative redshift distributions. We used both a Kolmogorov-Smirnov (KS) test and Anderson-Darling (AD) test with the null hypothesis that the EP and long GRB redshift distributions are drawn from the same underlying distribution. We have adopted the use of two non-parametric tests specifically because the KS test is most sensitive at the center (near the median) of the distributions, whereas the AD test is efficient at capturing deviations in the tails of the distributions. Therefore, running both tests allows us to probe deviations across the entire distribution.

We performed two-sample KS and AD tests, allowing us to compare two individual samples, i.e., all EP redshifts versus all long GRB redshifts. We made use of the multi-sample KS test function \texttt{ks\_2samp} within \texttt{SciPy} \citep{scipy}. For the two-sample AD test, the \texttt{SciPy} implementation (\texttt{anderson\_ksamp}) sets a built in $p$-value limit of 0.25, and, while it does not change our results, we instead opted to use an implementation within the R programming language\footnote{\url{https://search.r-project.org/CRAN/refmans/kSamples/html/ad.test.html}} and wrapped in \texttt{Python}\footnote{\url{https://rpy2.github.io/doc/latest/html/index.html}}. This choice was largely because the $p$-values for the AD test exceed the cap of $p_\textrm{AD}$\,$<$\,$0.25$ implemented in \texttt{SciPy}. We likewise tested a two-sample Cram\'er-von Mises test (\texttt{cramervonmises\_2samp} in \texttt{SciPy}), which provided similar results that did not change our conclusions. Therefore, for simplicity we focus on the KS and AD test results.

We consider a $p$-value of $<$\,$0.05$ as the boundary below which we reject the null hypothesis, which would suggest the distributions are different and the samples are unrelated. Instead, we find KS $p$-values in excess of this threshold ($p_\textrm{KS}$\,$\gg$\,$0.05$), supporting the null hypothesis and implying the distributions are derived from the same underlying distribution. We find KS $p$-values of $p_\textrm{KS}$\,$=$\,$0.43$, $0.34$, and $0.45$ for the comparison between long duration GRB redshifts and all EP sources, EP-GRBs, and EP sources without gamma-ray detections, respectively. In all cases, the $p$-value supports the null hypothesis. The same conclusion is drawn from an AD test, where we find $p_\textrm{AD}$\,$=$\,$0.47$, $0.24$, and $0.33$ for all EP sources, EP-GRBs, and EP sources without gamma-ray detections, respectively. These values are compiled in Table \ref{tab:stat_tests}.

\begin{table}[ht]
\centering
\caption{A compilation of the $p$-values derived in this work. 
}
\label{tab:stat_tests}
\begin{tabular}{lcc}
\hline\hline
\textbf{Sample} & $\mathbf{p}_\textbf{\textrm{KS}}$ & $\mathbf{p}_\textbf{\textrm{AD}}$ \\
\hline
All EP & 0.43 & 0.47 \\
EP-GRBs & 0.34 & 0.23 \\
No Gamma-rays & 0.45 & 0.33 \\
\hline
\end{tabular}
\tablecomments{All values refer to KS or AD tests where the sample of EP transient redshifts is compared to the population of long duration GRB redshifts (Figure \ref{fig:redshift-CDFs}; left panel).}
\end{table}

\subsubsection{Bootstrap Analysis}

In addition, we performed a bootstrap analysis by drawing $N$\,$=$\,$10,000$ samples and repeating the KS and AD calculation for each sample. We then derived the cumulative distribution of $p$-values for both statistics. Here, we briefly outline the procedure for the sample of all EP redshifts, but note that the same procedure is drawn for the three samples we consider here (all EP, EP-GRBs, and non-GRB EP sources). The procedure is as follows: \textit{i}) for a given sample of EP redshifts we produce a bootstrapped dataset of the same size (allowing for repetition), \textit{ii}) we compute the KS and AD statistic (comparing to the long GRB redshift distribution) using the bootstrapped sample, and \textit{iii}) we then repeat this procedure for $N$\,$=$\,$10,000$ bootstrapped samples. After this, \textit{iv}) we compute the 90\% confidence interval (CI) of the sample of CDFs, which are shown in the left panel of Figure \ref{fig:bootstrapKS}, and \textit{v}) we plot the cumulative distribution of $p$-values, shown in the right panel of Figure \ref{fig:bootstrapKS}. 
We likewise tested allowing for a two sample bootstrapping approach, where we also produced bootstrapped samples of long GRB redshifts prior to computing the two-sample test statistic. As the long GRB redshift distribution is sufficiently large, this does not have any impact and our result is the same in either case. This is shown in Figure \ref{fig:bootstrapKS}, where it can be seen that they converge to the same cumulative distribution of $p$-values.

Due to the small sample size of EP events, the distribution of KS $p$-values from bootstrapping reveals that $\sim$\,$30\%$ of the bootstrapped sample differs from the long GRB redshift distribution (i.e., $p_\textrm{KS}$\,$<$\,$0.05$; see Figure \ref{fig:bootstrapKS}). For the bootstrapped Anderson-Darling test we, find similar fractions of $\sim$\,$20$\,$-$\,$30\%$ with $p_\textrm{AD}$\,$<$\,$0.05$ (see Figure \ref{fig:bootstrapAD}). If the EP redshift distribution is indeed sampled from the long GRB distribution then increasing the population of EP events with measured redshifts will reveal this more robustly. At present, there is no strong statistical evidence to conclude they are drawn from separate distributions.

As an example of the strength of this conclusion, we performed the same statistical tests to compare the EP redshifts to those of short duration gamma-ray bursts \citep{OConnor2022}. The cumulative distribution of their redshifts is also shown in Figure \ref{fig:redshift-CDFs} versus long GRBs and EP redshifts. They are clearly visually distinct. A KS test yields a very convincing rejection of the null hypothesis with $p_\textrm{KS}$\,$=$\,$1.1\times10^{-4}$. We also performed the bootstrap analysis where we allow both the EP redshifts and short GRB redshifts to be resampled with repetition. We note that our result does not change if only one distribution is resampled. Through this bootstrap analysis, we find that 99.2\% of samples have a $p_\textrm{KS}$\,$<$\,$0.05$. This demonstrates that EP transients (at least the majority with redshifts) are unrelated to the compact object mergers that produce short duration GRBs.

\section{Discussion}
\label{sec:discussion}

\subsection{The Redshift Distribution of Einstein Probe Transients: Relation to Gamma-ray Bursts}
\label{sec:redshiftdist}

The redshift distribution of extragalactic transients provides useful information on their progenitors, and can be used to constrain their formation pathways. Here, we compare the redshift distribution of EP transients to both short and long gamma-ray bursts (see Figure \ref{fig:redshift-CDFs}). Due to broad range of redshifts over which EP transients are discovered (out to $z$\,$\sim$\,$5$) we cannot directly compare them to the distributions for optically discovered transients like supernovae. However, gamma-ray bursts are detected over this redshift range, extending even further back in the Universe's history.

To perform this comparison, we compiled all available redshifts of EP transients from the literature and from GCN Circulars (Table \ref{tab:EPtab}; \S \ref{sec:redshifts}). 
In total (through 2025-08-29), we identify 26 redshifts for EP transients, 12 of which have joint GRB detections. 
This comprises $\sim$\,$23\%$ percent of publicly reported (extragalactic) EP transients ($\sim$\,$113$ in total). The cumulative redshift distribution is shown in Figure \ref{fig:redshift-CDFs} for all EP transients (left panel) and the two sub-populations of those with and without GRB associations (right panel).

In Figure \ref{fig:redshift-CDFs}, we compare these distributions to the short duration GRB redshift distribution \citep{OConnor2022}, and the long duration GRB distribution compiled by the Greiner GRB Catalog. The short GRB redshift distribution is obviously inconsistent with a significant deviation due to the fact short GRBs are largely undetected\footnote{We note that this is likely a selection effect based on the difficulty to detect and then obtain a redshift for their host galaxies at higher redshifts, see, e.g., \citet{OConnor2022}.} at $z$\,$>$\,$2$, whereas half of all EP transients lie in that higher redshift range. Instead, the EP redshift distribution very clearly traces the long GRB distribution, signifying that they are likely produced by a similar set of progenitors (massive star collapse), or, at the very least, progenitors that have the same formation rate relative to the star formation history of the Universe \citep[see, e.g.,][]{Ghirlanda2022}. A Kolmogorov-Smirnov and Anderson-Darling test (see \S \ref{sec:bootstrap}; Figures \ref{fig:bootstrapKS} and \ref{fig:bootstrapAD}) favors that they are drawn from the same underlying distribution with a $p$-value of $\sim$\,$0.43$ (Table \ref{tab:stat_tests}). 

We note that it is known that there is at least some contamination within the population of EP/WXT transients from other classes of events (e.g., stellar flares, and X-ray binaries). However, we are confident that a least the extragalactic EP events with measured redshifts and GRB-like (X-ray and optical) afterglows are not coming from these other populations (see \S \ref{sec:EP}) and, therefore, they do not impact our conclusions. We further compared the distribution of EP transients with and without redshifts in the prompt ($0.5$\,$-$\,$4$ keV) X-ray flux and X-ray photon index plane ($F_\textrm{X,WXT}-\Gamma_\textrm{WXT}$) and verified that the distributions fully overlap with no discernible differences (i.e., both distributions span the same range of photon index and X-ray flux). Therefore, we do not expect selection biases within the extragalactic population of EP transients to play a large role in our conclusions.

While the sample of EP transients is significantly smaller than the hundreds of long GRBs with known redshifts \citep[e.g.,][]{Hjorth2012,Jakobsson2012,DAvanzo2014,Perley2016}, the current redshift distribution clearly favors a causal connection between \textit{Einstein Probe} detected fast X-ray transients and long duration gamma-ray bursts. This is further supported by their overlap in the standard $E_\textrm{p}$-$E_\textrm{iso}$ plane of gamma-ray bursts (Figure \ref{fig:Ep-Eiso}).
While the current sample of EP transients with reported peak energy $E_\textrm{p}$ constraints is notably small, and further analysis will be warranted upon the publication of a full catalog of EP transients, we note that the fact events like EP240315a and EP240801a follow the $E_\textrm{p}$-$E_\textrm{iso}$ strongly supports the hypothesis that EP events should follow these correlations. For example, these two events were missed by standard on-board GRB searches and would not have been detected (or identified) as GRBs without the temporal and spatial localization by EP allowing a targeted subthreshold search. It is therefore strongly implied that many other EP transients could be missed by typical gamma-ray monitors, even in a subthreshold search.

The EP mission is still relatively new (launched in January 2024), and it will continue to detect  high-energy transients in large numbers ($\sim$\,70 yr$^{-1}$). As a larger sample of events is built up over the course of the mission, we will eventually be able to separate more clearly into other subclasses of events, such as XRFs, EP-GRBs, and those with ``weak'' or ``failed'' jets (e.g., EP250108a; \citealt{Li2025}). This will allow for a robust comparison between the redshift and $E_\textrm{p}$-$E_\textrm{iso}$ distributions, and progenitors, of these subclasses (see \S \ref{sec:subclasses}). The specific breakdown of extragalactic EP transients into these subclasses, and their relative volumetric rates, is critical information on the deaths of massive stars and their ability to launch collimated relativistic jets \citep[see, e.g.,][]{Li2025}.

\subsection{The Separate Subclasses of Extragalactic Einstein Probe Transients: Are They All Collapsars?}
\label{sec:subclasses}

Since the 1990s, gamma-ray bursts have been subdivided based on their hardness ratios and the fraction of energy released at X-ray versus gamma-ray wavelengths. These subdivisions include X-ray flashes (XRFs) and X-ray rich GRBs \citep{Heise2001,Barraud2003,Sakamoto2005}. The narrow soft X-ray band of EP/WXT (without including additional higher energy instruments) does not allow for the historical hardness ratio definitions \citep[e.g.,][]{Sakamoto2004,Sakamoto2005,Sakamoto2008} to be applied. However, for a handful of EP-GRBs these definitions can be applied and confirm the transient as an XRF\footnote{It should be noted that standard XRF definitions rely on the observed peak energy (e.g., $<$\,$20$ keV; \citealt{Pelangeon2008}), which is not the same as an intrinsic XRF where the intrinsic (redshift corrected) peak energy is $<$\,$20$ keV \citep[e.g.,][]{Sakamoto2004}. We note that the majority of HETE-2 events with measured redshifts are not intrinsic XRFs (Figure \ref{fig:Ep-Eiso}).} \citep[e.g., EP240801a;][]{Jiang2025}. The subthreshold gamma-ray detection of many EP-GRBs \citep{Yin2024,Liu2024,Jiang2025} suggests that similar events have been missed historically, i.e., that they would not have been uncovered without EP's sensitive soft X-ray capabilities providing the detection necessary to allow a targeted search. This implies a strong selection effect against identifying fast X-ray transients as GRBs or XRFs, which may simply be misleading us as observers to conclude they are sources with different properties. Those differences aside, XRFs have generally been considered a natural extension of the GRB phenomena \citep[e.g.,][]{Heise2001}.

As an added complication, during the era of \textit{BATSE}, \textit{BeppoSAX}, and \textit{HETE-2} in the late 1990s and early 2000s, there was a more limited number of precise afterglow localizations, resulting in only a handful of redshifts\footnote{A literature search revealed only $\sim$20 redshifts for HETE-2 events overall with only 6 (observed) XRFs having a measured redshift \citep{Sakamoto2005,Pelangeon2008}.} for this class of events \citep{Barraud2003,Sakamoto2008,Pelangeon2008}. Despite the limited number of redshifts, the peak energy and fluence correlations were shown to be a natural extension of GRBs \citep{Sakamoto2008}, even without calculating their full energy release (Figure \ref{fig:Ep-Eiso}). This holds for recent EP-GRBs as well, see Figure \ref{fig:Ep-Eiso}. In this work, we have compared a significant sample of 26 redshifts for EP transients to long GRBs, demonstrating a secure connection between the two classes of events. The close match in their observed redshift distributions supports a shared progenitor population. However, such agreement is not necessarily expected\footnote{We note that there is a difference between an observed versus intrinsic redshift distribution, where the observed distribution depends heavily on the minimum observable fluence and whether a sample is volume limited.}, even under the assumption that both samples originate from the same progenitor, due to the likelihood of differing selection effects between EP/WXT and traditional high-energy monitors\footnote{For example, see Figure 14 of \citet{Pelangeon2008}, which shows the HETE-2 redshift distribution deviates from the early \textit{Swift} redshift sample.}. These possible selection biases may account for a slight discrepancies between the distributions that could become more evident with a larger EP sample.

For instance, it has been proposed that EP events should preferentially detect GRBs that are off-axis \citep{Gao2025}, mildly relativistic \citep[e.g.,][]{Busmann2025}, or at higher redshift \citep{Wei2025}, all of which are predicted to have lower peak energies in the observer frame. However, at present there is no strong evidence that EP preferentially detects higher redshift events at a higher rate than standard gamma-ray monitors. %Though we note that there is a slight deviation in the cumulative distribution towards higher redshifts at $z$\,$>$\,$2$ (Figure \ref{fig:redshift-CDFs}), but this may be due to the small number of events in our sample and potentially even out with time. We do note that it is unlikely that redshift is the dominant effect on the observed peak energy, as the scatter in peak energy is a natural property of GRBs (Figure \ref{fig:Ep-Eiso}). 
While this could explain subtle shifts in the redshift distribution, such as the mild excess of high redshift events at $z$\,$>$\,$2$ (Figure \ref{fig:redshift-CDFs}), the current evidence is insufficient to conclude that EP detects higher redshift events at an elevated rate. This observed deviation may be a statistical fluctuation given the modest sample size and could even out with time. Moreover, it is unlikely that redshift is the dominant factor in determining the observed peak energy, as significant scatter is an intrinsic property of GRBs (Figure \ref{fig:Ep-Eiso}).

Despite this close connection to long GRBs, EP has uncovered a variety of peculiar transients that are missed by traditional gamma-energy monitors (Figure \ref{fig:lum}). These include candidate relativistic jetted tidal disruption events (EP240408a, \citealt{OConnor2025,Zhang2025}; and EP250702a/GRB 250702B, \citealt{,OConnor2025ep,Levan2025,Neights2025,Carney2025,Gompertz2025}) and multiple events with extreme optical rebrightening episodes \citep[EP240414a and EP241021a;][]{vanDalen2024,Busmann2025}. The physical origin of these rapid ($\sim$\,$1$ d) and large amplitude ($\sim$\,$1$ mag) rebrightening episodes are debated. While it has been proposed that discrete refreshed shocks provide the most straightforward explanation \citep{Srivastav2024,Busmann2025}, alternative models suggest that the initially fading emission is due to a cocoon produced by the interaction of the jet with an extended stellar envelope \citep{vanDalen2024,Hamidani2025,Zheng2025,Gianfagna2025}. In the latter case, the jet could either be mildly relativistic or an ultrarelativistic outflow viewed off-axis \citep{Zheng2025,Yadav2025}.

EP has also revealed rare transient classes with unprecedented frequency. For example, it has already detected two relativistic shock breakout candidates \citep[EP250108a and EP250304a;][]{Rastinejad2025EP,Eyles-Ferris2025EP,Srinivasaragavan2025EP0108a,EP250304a-SN-GCN}, already exceeding the (observed) rate in the past 20 years from \textit{Swift} \citep{Soderberg2006grb060218,Campana060218}. These are among the few EP events (all lacking gamma-rays and at $z$\,$<$\,$0.4$; Figure \ref{fig:lum}) associated to Type Ic-BL supernovae; the same supernovae generally associated to long duration GRBs and their collapsar progenitors \citep{Woosley2006,Hjorth2012sn}. That these low redshift EP transients show clear supernova signatures is consistent with our interpretation that many EP transients originate from the deaths of massive stars. In the case of EP250108a, it has also been suggested that the jet potentially failed to breakout of its stellar envelope due to an extended circumstellar envelope \citep{Eyles-Ferris2025EP,Srinivasaragavan2025EP0108a}. 
Therefore, while EP is very likely identifying  events that are driven by the same progenitors as long GRBs (tracing the same redshift distribution that closely follows star formation), these events also cover a parameter space that is not probed by standard gamma-ray-only triggers and are likely the predicted missing population of failed jets and dirty fireballs \citep{Rhoads2003} with lower bulk Lorentz factors.

Nevertheless, the EP sample is unlikely to be composed solely of collapsars. While we have filtered out Galactic contaminants, extragalactic events such as EP250704a fall within the region typically occupied by short GRBs in the
%Despite this conclusion, we must note that the extragalactic population of EP transients compiled here, while free of contamination from Galactic impostors, is unlikely to be a completely pure sample of collapsars. For example, EP250704a is clearly falls with the location of short duration GRBs in the 
$E_\textrm{p}$-$E_\textrm{iso}$ plane (see Figure \ref{fig:Ep-Eiso}; \citealt{250704a-grb}), and may be associated to compact object mergers. 
%However, the existence of short-duration collapsars \citep{Ahumada2021} complicates the classification and late-time observations are necessary to exclude the existence of a supernova component. 
Importantly, the inclusion or exclusion of such events has negligible impact on our statistical conclusions. %In any case, we expect this possible contamination from non-collapsars to be small enough that it does not impact our conclusions, and that the majority of events are still produced by jet's launched through stellar death. 
For example, removing EP250704a from our sample results in only a marginal change in $p$-values (remaining at $\approx$\,$0.3$) from both KS and AD tests, and does not modify our conclusion (see \S \ref{sec:bootstrap}). %As a larger sample of events is compiled, the potential impurities in the sample will become more clear (e.g., short versus long GRBs), and their contamination to the bulk population even less impactful.
As the sample grows, the identification and influence of potential contaminants (e.g., short GRBs) will become clearer and less significant to the overall population.

%\textcolor{purple}{also add comment on if XRFs are choked jets they are likely wider so beaming corrections are necessary.}
The volumetric rates of EP transients remain poorly constrained, but preliminary estimates suggest that they may be comparable to or exceed those of standard GRBs. For EP240414a, \citet{Sun2024} set a lower limit of $0.3$ Gpc$^{-3}$ yr$^{-1}$ to its volumetric rate. For EP250108a, \citet{Li2025} derived $7.3^{+16.8}_{-6.0}$ Gpc$^{-3}$ yr$^{-1}$. After correcting for the redshift completeness of EP transients, they find a corrected rate of $29.2^{+67.2}_{-24.0}$ Gpc$^{-3}$ yr$^{-1}$ \citep{Li2025}. This is comparable to the volumetric rate of low luminosity GRBs ($\sim$\,$100$ Gpc$^{-3}$ yr$^{-1}$) like GRB 060218 \citep{Soderberg2006grb060218}, and significantly exceeds that of typical high-luminosity GRBs ($1.3^{+0.6}_{-0.7}$ Gpc$^{-3}$ yr$^{-1}$; \citealt{Wanderman2010}) - where for both low and high luminosity GRBs the rates above correspond to events that are beamed towards Earth. 
This supports a scenario in which weak or failed jets are intrinsically more common than successful ones, and that prior selection biases against soft X-ray transients hindered their discovery. As the \textit{Einstein Probe} continues to expand the sample of such events, our understanding of their origins, diversity, and true rates will improve significantly.

\section{Conclusions}
\label{sec:conclusions}

In this work, we compared the cumulative redshift distributions of \textit{Einstein Probe} fast X-ray transients and long-duration gamma-ray bursts using non-parametric two-sample tests. These tests reveal that the distributions are statistically indistinguishable, supporting that their redshift distributions are drawn from the same underlying population. Therefore, the fraction of EP transients related to GRBs is higher than would be expected based on gamma-ray associations alone. We conclude that a substantial fraction of EP transients, at the very least those with measured redshifts, originate from similar massive star progenitors (collapsars) as long duration gamma-ray bursts. This is supported by the similar redshift distribution for EP transients both with and without prompt gamma-ray detections, as well as their continuity with long GRBs in the $E_\textrm{p}$-$E_\textrm{iso}$ plane.

Several mechanisms can suppress a gamma-ray detection from otherwise GRB-like explosions (e.g., weak or trapped jets, slightly off-axis viewing angles, lower bulk Lorentz factors, or limited instrument sensitivity). The low-$z$ ($z<0.4$) subsample, in particular, is consistent with prompt spectra peaking in the soft X-ray band, naturally explaining the absence of high-energy triggers. While small number statistics and selection effects remain, our redshift and prompt emission comparisons support a predominantly massive star origin for \textit{Einstein Probe} fast X-ray transients. Our results reinforce similar conclusions drawn between XRFs and GRBs based on \textit{BATSE}, \textit{BeppoSAX}, and \textit{HETE-2}.

%% Please use the acknowledgment and contribution environments. This will 
%% be anonomyized when the "anonymous" style option is used. 
\begin{acknowledgments}

The authors acknowledge the referee for their careful review of the manuscript. 
The authors thank the \textit{Einstein Probe} team for their public release of on-board WXT triggers that enables multi-wavelength follow-up. B. O. acknowledges useful discussions with Dheeraj Pasham, Igor Andreoni, Jamie Kennea, and Jimmy DeLaunay. M. B. acknowledges Fabian Schüssler for assistance with the Astro-COLIBRI platform. B. O. thanks Rosa Becerra and Massine El Kabir for their aid in obtaining X-shooter spectra.

B. O. is supported by the McWilliams Postdoctoral Fellowship in the McWilliams Center for Cosmology and Astrophysics at Carnegie Mellon University. M. B. is supported by a Student Grant from the Wübben Stiftung Wissenschaft. This work benefited from travel support for M. B. from the McWilliams Visitors Program as part of the McWilliams Center for Cosmology and Astrophysics at Carnegie Mellon University. P. B. is supported by a grant (no. 2020747) from the United States-Israel Binational Science Foundation (BSF), Jerusalem, Israel, by a grant (no. 1649/23) from the Israel Science Foundation and by a grant (no. 80NSSC 24K0770) from the NASA astrophysics theory program. E. T., N. P., and Y. Y. are supported by the European Research Council through the Consolidator grant BHianca (grant agreement ID 101002761). 

Based on observations obtained at the international Gemini Observatory, a program of NSF's OIR Lab, which is managed by the Association of Universities for Research in Astronomy (AURA) under a cooperative agreement with the National Science Foundation on behalf of the Gemini Observatory partnership: the National Science Foundation (United States), National Research Council (Canada), Agencia Nacional de Investigaci\'{o}n y Desarrollo (Chile), Ministerio de Ciencia, Tecnolog\'{i}a e Innovaci\'{o}n (Argentina), Minist\'{e}rio da Ci\^{e}ncia, Tecnologia, Inova\c{c}\~{o}es e Comunica\c{c}\~{o}es (Brazil), and Korea Astronomy and Space Science Institute (Republic of Korea). The data were acquired through the Gemini Observatory Archive at NSF NOIRLab and processed using DRAGONS (Data Reduction for Astronomy from Gemini Observatory North and South). Based on observations collected at the European Organisation for Astronomical Research in the Southern Hemisphere under ESO programme(s) 114.27LW. 

\end{acknowledgments}

%\begin{contribution}
%%This section gives authors the space to recognize author contributions. The text inside this environment is NOT counted towards the total word quanta. At a minimum, manuscripts are expected to include this text:

%...

%% But authors are expected to provide more specific details, e.g. 
%%
%%SC was responsible for writing and submitting the manuscript.
%%WWM came up with the initial research concept and edited the manuscript.
%%OTS obtained the funding and edited the manuscript.
%%EBF provided the formal analysis and validation. He also edited the manuscript.
%%GEH Supervised the undergraduates, wrote the software and administers the project github and Zenodo repositories.
%%
%% Authors can use the Contributor Role Taxonomy (CRediT) at
%% for ideas on how write a good statement tailored to their needs.

%\end{contribution}

\facilities{\textit{Einstein Probe}, Gemini-South Telescope, Very Large Telescope
}

\software{\texttt{Astropy} \citep{2018AJ....156..123A,2022ApJ...935..167A}, \texttt{SciPy} \citep{scipy}, \texttt{DRAGONS} \citep{Labrie2019,Labrie2023}, X-shooter pipeline \citep{Goldoni2011}, \texttt{zHunter} \citep{Palmerio_zHunter_2025}
}


\appendix

\section{Optical Spectroscopy with Gemini and the VLT}
\label{sec:optspec}

Here we report on three spectroscopic redshifts of EP transients obtained through our programs on the Gemini-South Telescope and the Very Large Telescope (VLT). The spectra are shown in Figure \ref{fig:speczs}.

\subsection{EP250302a}
\label{sec:250302a}

We carried out ultraviolet to near-infrared spectroscopy of EP250302a with the Very Large Telescope located at Cerro Paranal, Chile using the X-shooter spectroscopy \citep{Vernet2011} under program 114.27LW (PI: Troja). Observations with the X-shooter spectrograph mounted on the ESO VLT UT3 (Melipal) began on 2025-03-03 at 05:14:45 UT, corresponding to $T_0+13.6$ hr, for $4\times600$ s. The data were reduced using the standard X-shooter pipeline \citep{Goldoni2011} and revealed a trace in all three arms (UVB, VIS, NIR).In the UVB arm, we identify multiple narrow absorption features associated with Fe\RN{2}$_{\lambda2600, 2587, 2383, 2374, 2344}$ at redshift $z=1.1310\pm 0.0002$. A zoom in on the spectrum showing these features is displayed in Figure \ref{fig:speczs}. This determination is consistent with earlier reports for the redshift \citep{0302a-vlt-redshift,2025GCN.39574....1Y}. We further identify an intervening absorber at $z=0.549$, as reported by  \citet{2025GCN.39561....1Y}. % based on the identification of narrow absorption features at XX, XX corresponding to XX, XX.} 
No other absorbers or unidentified absorption lines are discovered so we consider the redshift secure. As we do not identify any fine-structure absorption lines at $z=1.1310$, we cannot exclude a higher redshift. However, we detect a weak trace down to 370 nm, which sets an upper limit of $z$\,$<$\,$2.0$ to the redshift of EP250302a.

\subsection{EP250821a}
\label{sec:250821a}

We observed EP250821a with the Gemini GMOS-S spectrograph starting on 2025-08-26 at 01:53:43 UT under program GS-2025A-FT-111 (PI: O'Connor). Our spectra cover wavelengths 575 to 1060 nm and consist of 4$\times$1200 s exposures using the R400 grating. The data were reduced using the \texttt{DRAGONS} software \citep{Labrie2019,Labrie2023}. We detect multiple emission lines from the underlying host galaxy, including the [OII]$_{\lambda3727,3729}$ doublet, [OIII]$_{\lambda4960,5007}$, H$\beta$, and H$\alpha$, at a consistent redshift of $z=0.5775\pm 0.0003$ as initially reported by \citet{250821a-z} and \citet{250821a-z2}. We also detect [NaI] absorption and the 4000\AA\, break at this redshift. We note that \citet{250821a-z} reported the identification of multiple narrow absorption features in an earlier optical spectrum obtained $\sim$\,$4$ days prior. This supports our redshift determination from host galaxy emission lines (Figure \ref{fig:speczs}).

\subsection{EP250827a}
\label{sec:250827a}

\begin{figure*}
    \centering
\includegraphics[width=\columnwidth]{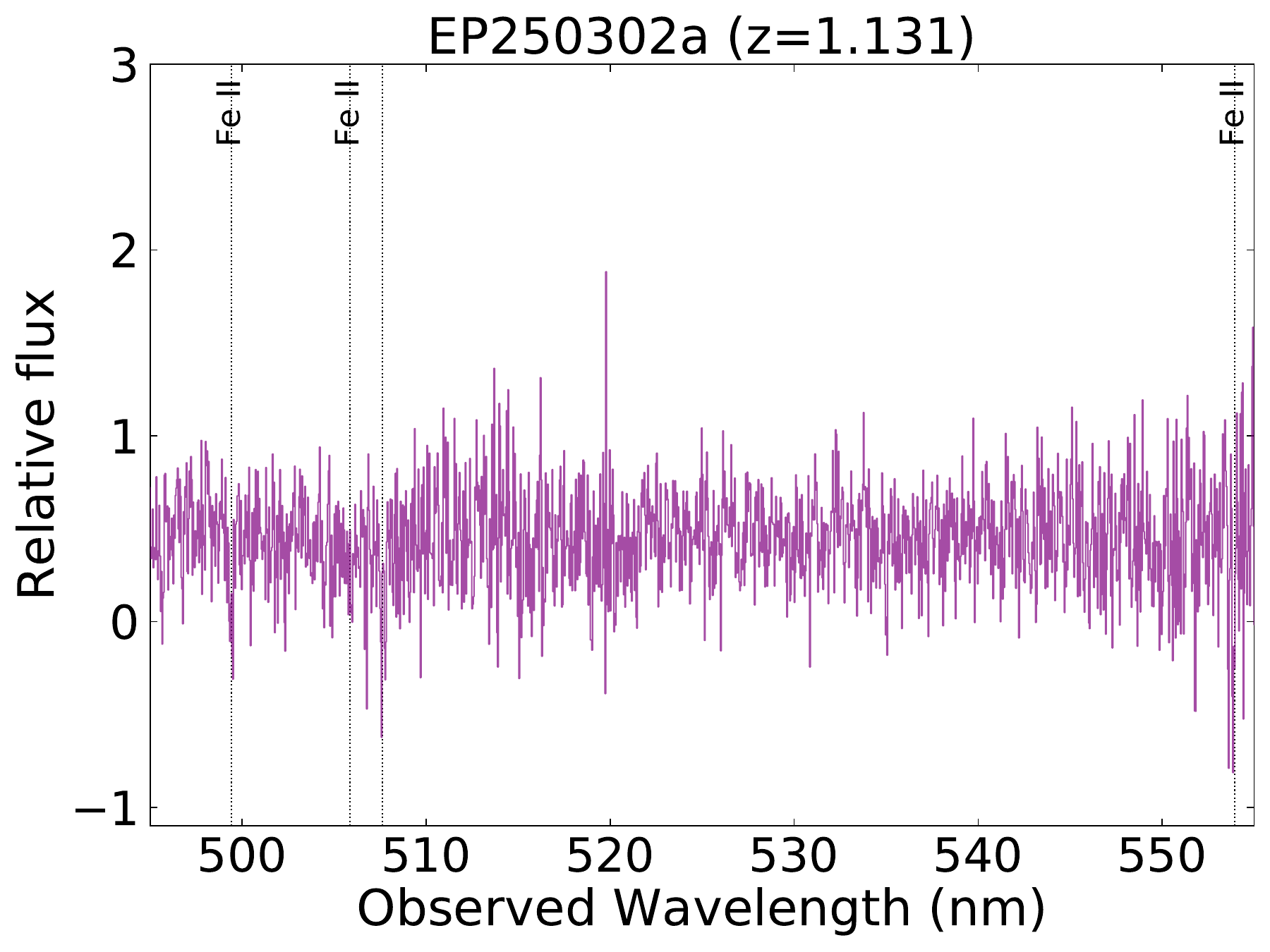}
\includegraphics[width=\columnwidth]{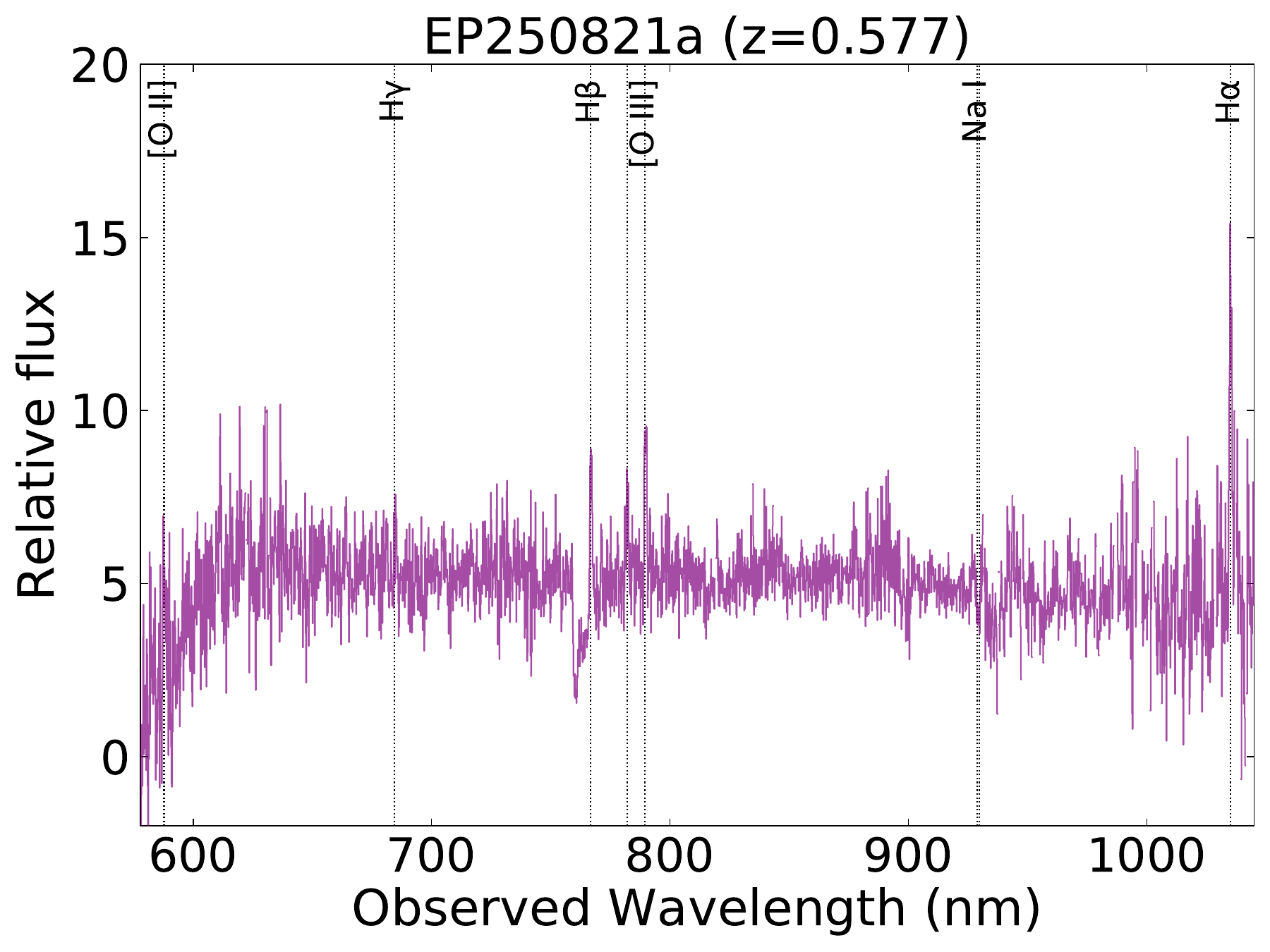}
\includegraphics[width=\columnwidth]{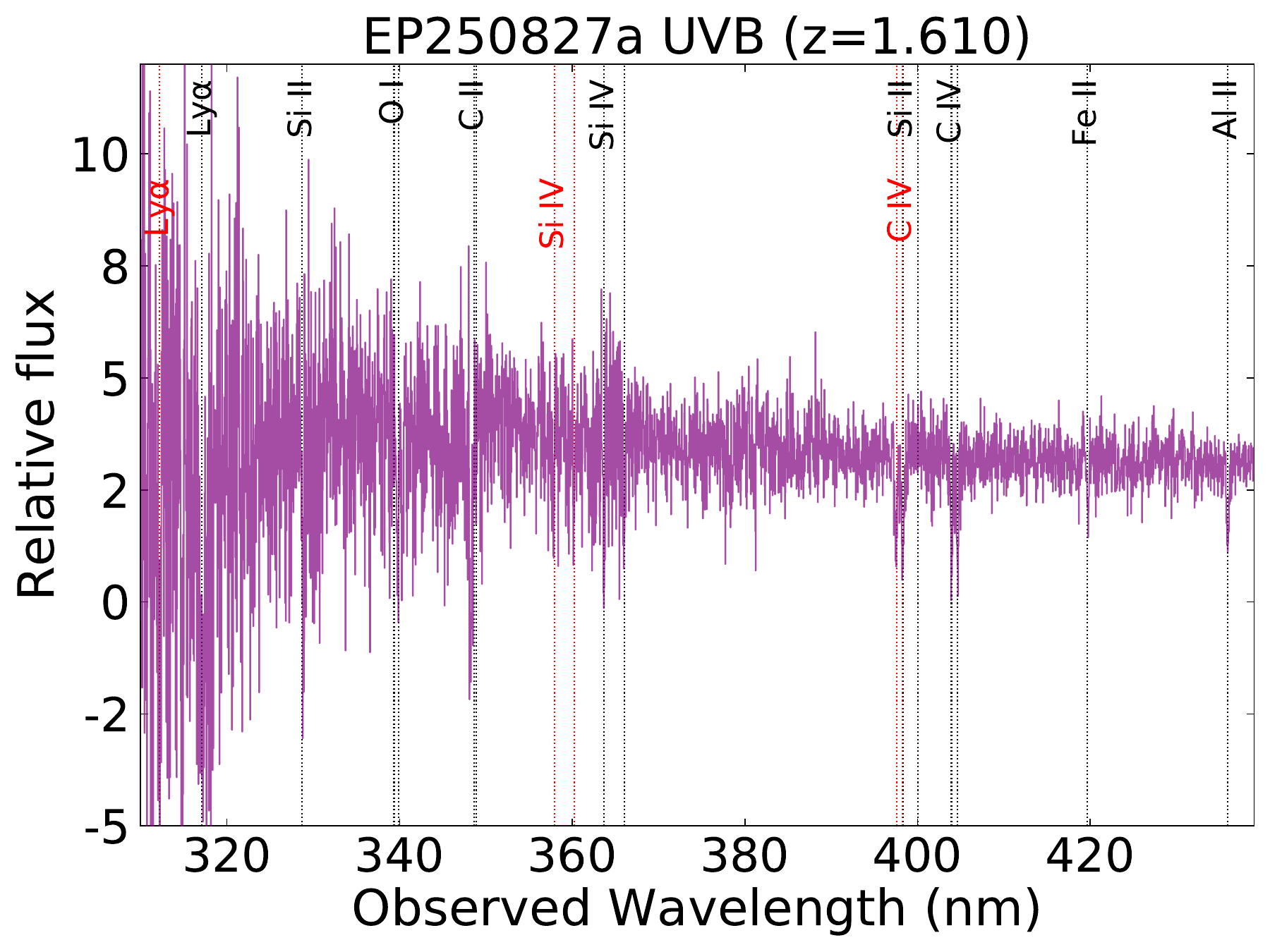}
\includegraphics[width=\columnwidth]{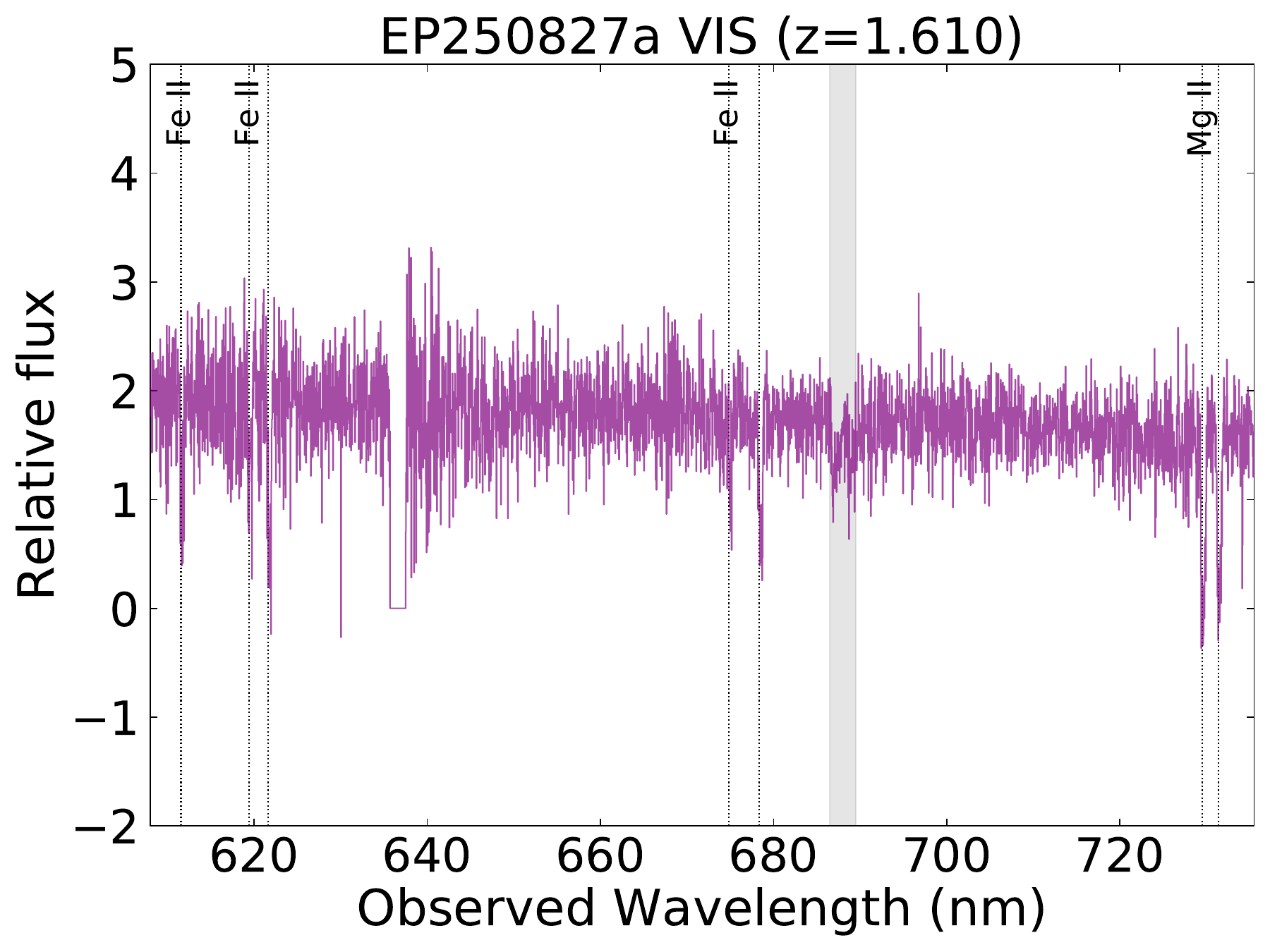}
    \caption{Spectroscopy of EP250302a (VLT X-shooter), EP250821a (Gemini-South GMOS), and EP250827a (VLT X-shooter) used to measure their redshifts. For EP250827a, due to the large number of absorption lines, we show both the UVB and VIS arms of the X-shooter spectrum. The spectra have not been smoothed or re-binned. 
    }
    \label{fig:speczs}
\end{figure*}

We observed GRB 250827A with the VLT X-shooter spectrograph under program 114.27LW (PI: Troja) starting on 2025-08-27 at 09:21:51 UT, corresponding to $T_0+1.67$ hr, for $4\times600$ s. The data were reduced using the standard X-shooter pipeline \citep{Goldoni2011}. Our spectrum reveals a large number of significant absorption features. We identify $>$\,$20$ narrow absorption lines between the UVB and VIS arms consisting of Si\RN{2}$_{\lambda1260,1304}$, O\RN{1}$_{\lambda1302}$, C\RN{2}$_{\lambda1334,1335}$, Si\RN{4}$_{\lambda1394,1403,1527,1533}$, C\RN{4}$_{\lambda1548,1550}$, Fe\RN{2}$_{\lambda1608}$, Al\RN{2}$_{\lambda1671}$, and Al\RN{3}$_{\lambda1855,1863}$ in the UVB arm and Fe\RN{2}$_{\lambda2344,2345,2382,2383,2384,2587,2600}$ and Mg\RN{2}$_{\lambda2796,2804}$ in the VIS arm. These features are all at a common redshift of $z=1.6105\pm 0.0003$. The existence of fine-structure lines secures the redshift of the GRB. We zoom in on a few of these features in Figure \ref{fig:speczs}. 
This redshift is consistent with previous reports by \citet{250827a-z} and \citet{250827a-z2}. We identify a marginal broad absorption feature at the blue end of the UVB arm that could be associated to Lyman $\alpha$ (Ly$\alpha$) at this redshift of $z=1.6105$. This is supported by the detection of a blue trace down to 320 nm, which sets an upper limit to the redshift of $z$\,$<$\,$1.63$. This is consistent with the lack of detection of any additional higher redshift absorption features. We do, however, identify an intervening absorber at $z=1.5688$ due to measuring narrow Si\RN{4}$_{\lambda1394,1403}$ and C\RN{4}$_{\lambda1548,1550}$ absorption lines. No other absorbers or unidentified absorption lines are discovered. As such, we consider the redshift secure.

\section{Comments on Redshifts Based on Host Galaxy Emission Lines}
\label{caution}

The transients with redshifts based on emission lines from their underlying host galaxies are: 
EP240414a \citep{vanDalen2024,Srivastav2024}, EP241107A \citep{241107a-z}, EP241113a \citep{241113a-z}, EP241217b/GRB 241217A \citep{241217b-z}, EP250416a/ GRB 250416C \citep{250416a-z}, and EP250821a \citep[\S \ref{sec:250821a};][]{250821a-z,250821a-z2}. We note that in the case of EP250821a (Figure \ref{fig:speczs}), \citet{250821a-z} also reported absorption features. In particular, the redshifts for EP241113a, EP241217b, and EP250416a are tentative as they are based on a single emission line that is interpreted as [OIII]$_{\lambda5007}$ for EP241217b \citep{241217b-z}, and the [OII]$_{\lambda3727,3729}$ doublet for EP241113a \citep{241113a-z} and EP250416a \citep{250416a-z}.
%EP241021a (but also absorption lines, which we confirmed based on the X-shooter spectrum) so do not include)
We note that the redshift of EP240414a is more secure as the galaxy's redshift matches the distance inferred from supernova features in a sequences of optical spectra \citep{vanDalen2024}. We have excluded EP250207B from our sample due to its uncertain redshift \citep[][R. L. Becerra et al., in prep.]{Jonker2025}, but note that the exclusion of this single event does not impact our conclusions.

\section{Bootstrap Analysis Results}

\begin{figure*}
    \centering
\includegraphics[width=1.05\columnwidth]{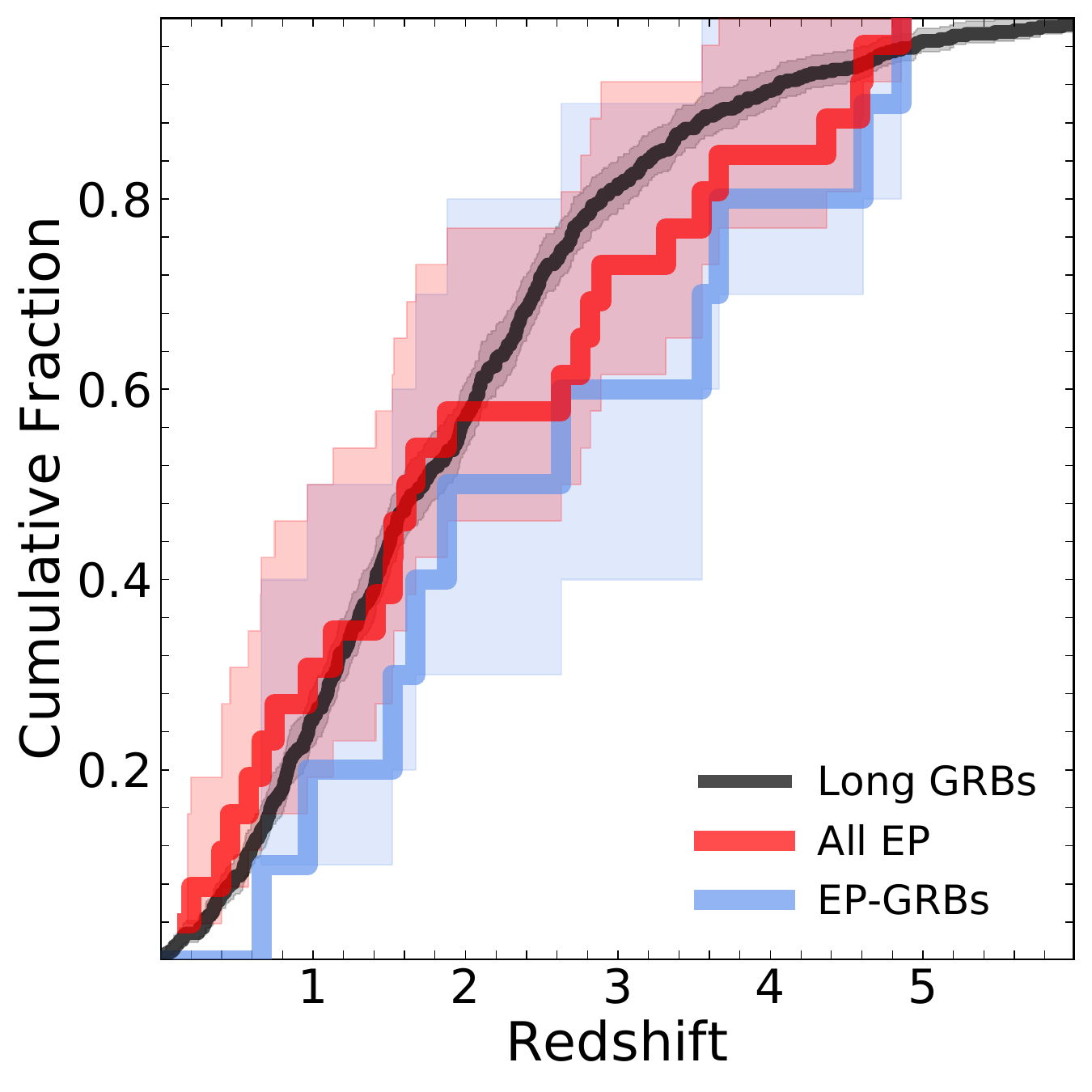}
\includegraphics[width=1.05\columnwidth]{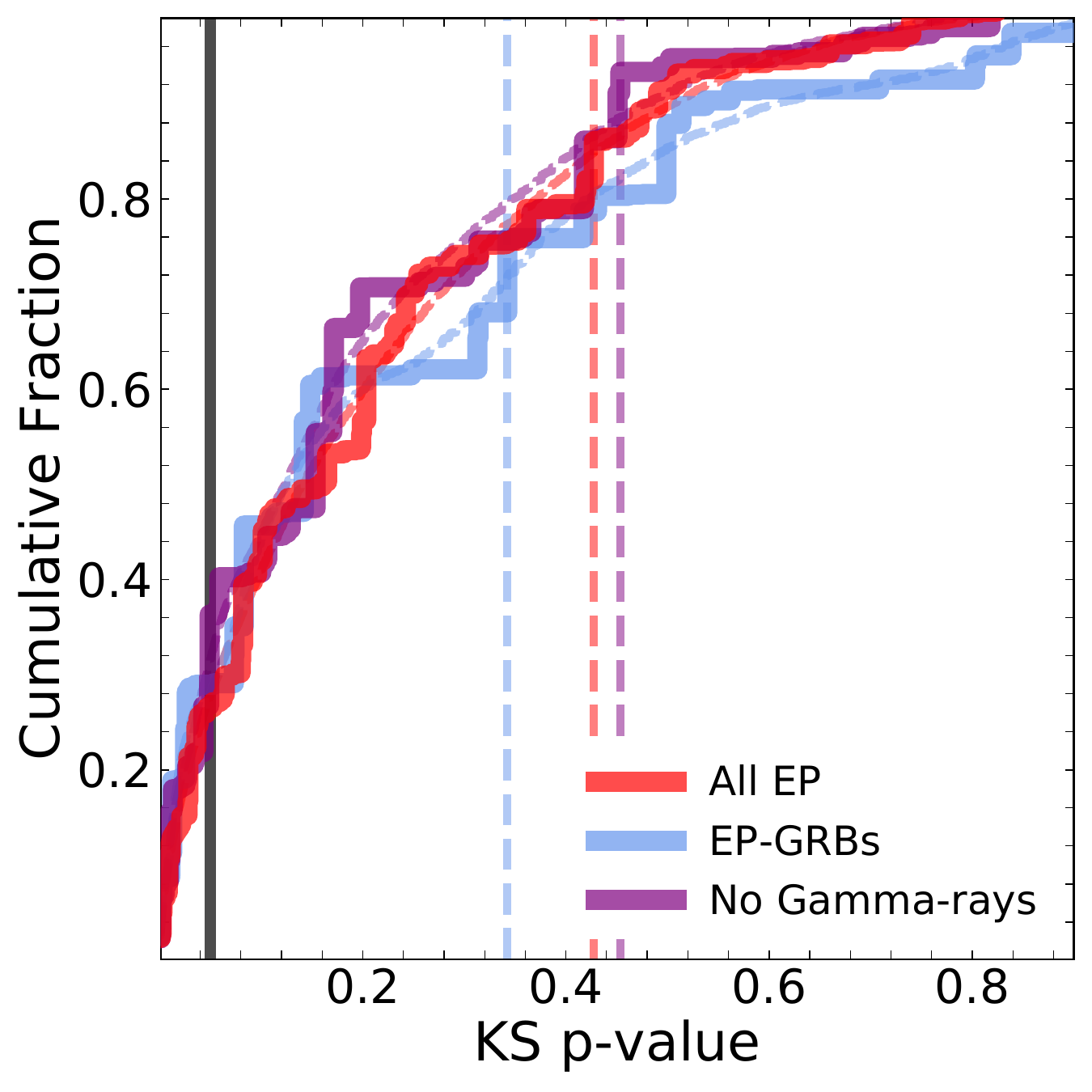}
    \caption{\textbf{Left:} Cumulative distribution of the redshift of EP transients (red) compared to EP-GRBs (blue) and long GRBs (black). The shaded regions show the 90\% confidence regions of the CDFs after bootstrapping $N$\,$=$\,$10,000$ times. To avoid further crowding the figure we do not display the EP sub-sample without gamma-rays. \textbf{Right:} Cumulative distribution of Kolmogorov-Smirnov $p$-values obtained from bootstrapping. We show the $p$-value distributions for all EP transients (red), EP-GRBs (blue), and those without prompt gamma-ray detections (purple). The thin line CDFs are computed when also bootstrapping the long GRB distribution, whereas the thick solid lines are determined when only bootstrapping the three EP distributions. The vertical lines show the $p$-value for the measured sample without any bootstrapping. The solid black line denotes $p$\,$=$\,$0.05$, below which the null hypothesis is rejected. 
    }
    \label{fig:bootstrapKS}
\end{figure*}

\begin{figure}
    \centering
\includegraphics[width=1.0\columnwidth]{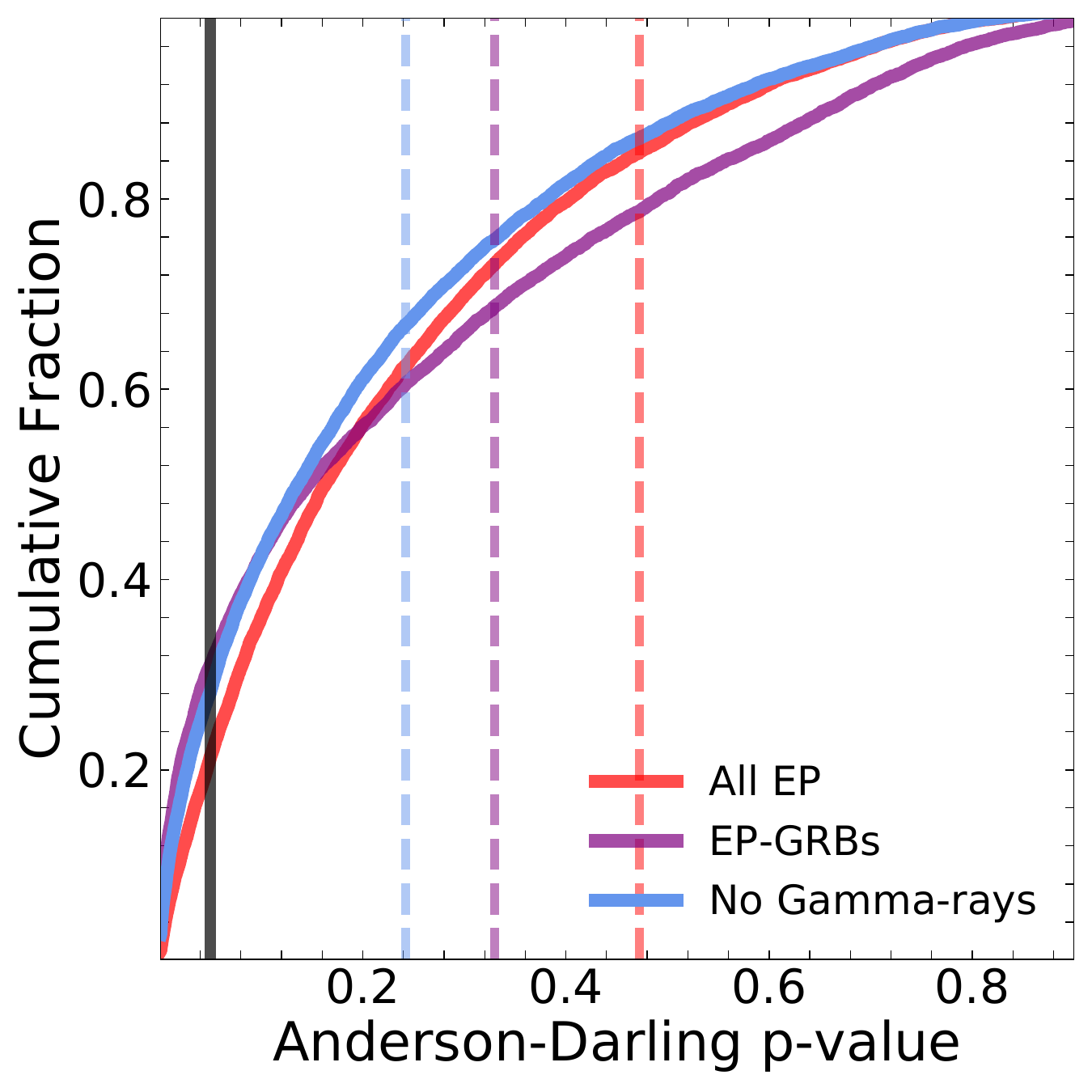}
    \caption{Cumulative distribution of Anderson-Darling $p$-values obtained from bootstrapping. We show the $p$-value distributions for all EP transients (red), EP-GRBs (blue), and those without prompt gamma-ray detections (purple). The vertical lines show the $p$-value for the measured sample without any bootstrapping. The solid black line denotes $p$\,$=$\,$0.05$, below which the null hypothesis is rejected. 
    }
    \label{fig:bootstrapAD}
\end{figure}

We performed a bootstrap analysis between the EP redshift distribution and the redshift distribution of long gamma-ray burst (see \S \ref{sec:bootstrap}). The cumulative distribution of bootstrapped $p$-values computed using the Kolmogorov-Smirnov test are displayed in Figure \ref{fig:bootstrapKS} and for the Anderson-Darling test the distributions are shown in Figure \ref{fig:bootstrapAD}. We find that $\sim$\,$30\%$ of bootstrapped samples yield a Kolmogorv-Smirnov test $p$-value $p_\textrm{KS}$\,$<$\,$0.05$ and $\sim$\,$20$\,$-$\,$30\%$ for the Anderson-Darling test with $p_\textrm{AD}$\,$<$\,$0.05$.

%\appendix

%\clearpage

\bibliography{bib}{}
\bibliographystyle{aasjournal}



\end{document}